\documentclass[sort&compress]{elsarticle}

\usepackage{amssymb}
\usepackage{todonotes}
\usepackage{glossaries}
\usepackage{graphicx}
\usepackage{siunitx}
\usepackage{caption}
\usepackage{subcaption}
\usepackage{marginnote}
\usepackage{booktabs}
\usepackage{circledsteps}
\usepackage{tikz}
\usepackage[singlespacing]{setspace}

\newacronym{acr:OFDR}{OFDR}{Optical Frequency Domain Reflectometry}
\newacronym{acr:OTDR}{OTDR}{Optical Time Domain Reflectrometry}
\newacronym{acr:FBG}{FBG}{Fiber Bragg Grating}
\newacronym{acr:NA}{NA}{Numerical Aperture}
\newacronym{acr:BMS}{BMS}{battery management system}
\newacronym{acr:LIB}{LIB}{lithium-ion battery}

\newcommand{\mi}[2]{\ensuremath{{#1}_{\mathrm{#2}}}}     			
\newcommand*\circled[1]{%
	\tikz[baseline=(char.base)]{
		\node[shape=circle,draw,inner sep=0pt, minimum size=1.5em, text height=1em, text depth=0.25em] (char) {\strut #1};
	}%
}

\journal{Energy Storage}

\begin{document}

	\begin{frontmatter}
			
		\title{Advancing Measurement Capabilities in Lithium-Ion Batteries: Exploring the Potential of Fiber Optic Sensors for Thermal Monitoring of Battery Cells}
		
		\author[label1,label2,label3]{Florian Krause\corref{cor1}}
		\ead{florian.Krause@isea.rwth-aachen.de}
		\ead[url]{https://www.isea.rwth-aachen.de}
		\author[label1,label2]{Felix Schweizer}
		\author[label5]{Alexandra Burger}
		\author[label1,label2]{Franziska Ludewig}
		\author[label1,label2,label3]{Marcus Knips}
		\author[label1,label2,label3]{Katharina Quade}
		\author[label5]{Andreas Würsig}
		\author[label1,label2,label3,label4]{Dirk Uwe Sauer}
		
		\affiliation[label1]{organization={Chair for Electrochemical Energy Conversion and Storage Systems, Institute for Power Electronics and Electrical Drives (ISEA), RWTH Aachen University}, 
			addressline={Campus-Boulevard 89}, 
			city={Aachen},            
			postcode={52074}, 
			country={Germany}}
		
		\affiliation[label2]{organization={Center for Ageing, Reliability and Lifetime Prediction of Electrochemical and Power Electronic Systems (CARL), RWTH Aachen University}, 
			addressline={Campus-Boulevard 89}, 
			city={Aachen},            
			postcode={52074}, 
			country={Germany}}
		
		\affiliation[label3]{organization={Juelich Aachen Research Alliance, JARA-Energy}, 
			addressline={Templergraben 55}, 
			city={Aachen},            
			postcode={52056}, 
			country={Germany}}
		
		\affiliation[label4]{organization={Helmholtz Institute Münster (HIMS), IEK 12, Forschungszentrum}, 
			city={Jülich},            
			postcode={52425},
			country={Germany}}
		
		\affiliation[label5]{organization={Fraunhofer Institute for Silicon Technology (ISIT)}, 
			addressline={Fraunhoferstraße 1}, 
			city={Itzehoe},            
			postcode={25524},
			country={Germany}}
		
		\cortext[cor1]{Corresponding author.}

		\begin{abstract}
			
			This work demonstrates the potential of fiber optic sensors for measuring thermal effects in lithium-ion batteries, using a fiber optic measurement method of \acrfull{acr:OFDR}. The innovative application of fiber sensors allows for spatially resolved temperature measurement, particularly emphasizing the importance of monitoring not just the exterior but also the internal conditions within battery cells. Utilizing inert glass fibers as sensors, which exhibit minimal sensitivity to electric fields, opens up new pathways for their implementation in a wide range of applications, such as battery monitoring. The sensors used in this work provide real-time information along the entire length of the fiber, unlike commonly used \acrfull{acr:FBG} sensors. It is shown that using the herein presented novel sensors in a temperature range of $0-\SI{80}{\degreeCelsius}$ reveals a linear thermal dependency with high sensitivity and a local resolution of a few centimeters. 
			Furthermore, this study presents preliminary findings on the potential application of fiber optic sensors in \acrfull{acr:LIB} cells, demonstrating that the steps required for battery integration do not impose any restrictive effects on thermal measurements.
			
		\end{abstract}
		
		\begin{graphicalabstract}
			\includegraphics[width=0.9\linewidth]{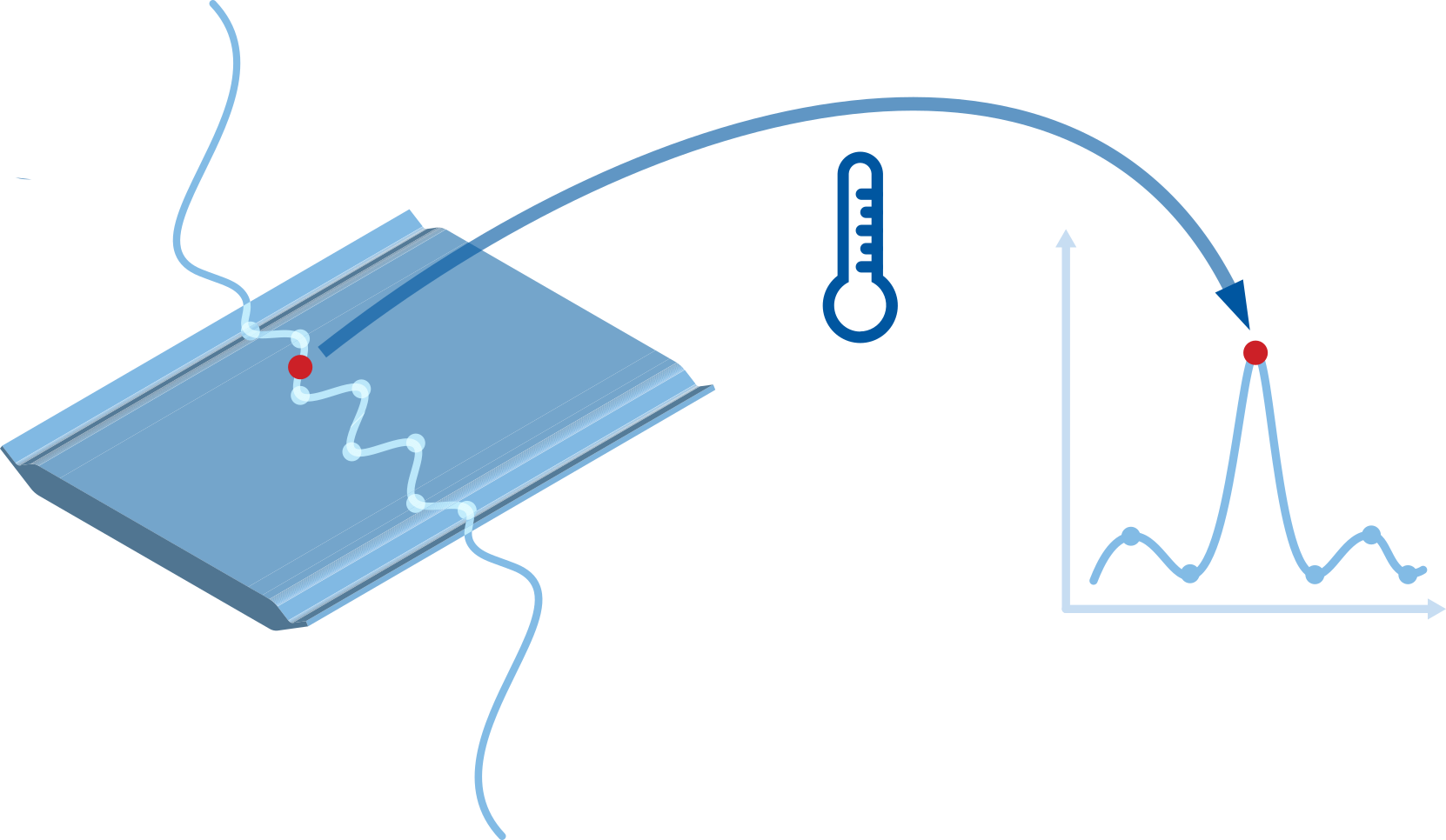}
		\end{graphicalabstract}

		\begin{highlights}
			\item Development and analysis of fiber optic sensors for precise thermal monitoring in lithium-ion battery cells.
			\item Demonstration of spatially resolved temperature measurements with high sensitivity and accuracy.
			\item Assessment of mechanical and thermal effects of fiber optic sensor integration on measurement reliability.
		\end{highlights}

		
	\end{frontmatter}
	
	
	\section{Introduction}
	\label{kap:Introduction}
	
	In recent years, battery storage systems have taken on an exceptional role in energy storage technology. As a result, lithium-ion batteries have become indispensable in the field due to their high energy and power density \cite{wegmann_Assessing_2018, figgener_development_2022, rothgang_Battery_2015}. With increasing energy contents of battery systems, enabling higher performance and longevity, safety requirements become an increasingly important aspect \cite{bausch_Naturallyderived_2023,li_30_2018}.
	However, present condition monitoring of \acrshort{acr:LIB}~s is limited to current, voltage and temperature measurements on the outside of the cell \cite{lelie_Battery_2018}. The optimum operating temperature of lithium-ion cells is usually specified as $\SI{15}{\degreeCelsius}$ to $\SI{35}{\degreeCelsius}$, but can also vary depending on the cell chemistry \cite{cloos_Challenges_2024, ma_Temperature_2018}. Both above and below this temperature range, temperature-dependent effects appear which significantly affect performance, capacity and aging of the battery \cite{waldmann_Temperature_2014, bandhauer_Critical_2011, paarmann_Measurement_2021}. Due to the impact of temperature on \acrshort{acr:LIB} performance, many studies have focused on the temperature distribution and its influence within the cell \cite{cloos_Thermal_2024, werner_Inhomogeneous_2020}, as well as between the cells of a battery pack \cite{docimo_Analysis_2019, altaf_Simultaneous_2014, pesaran_Thermal_1997}. \textit{Rahn et al.} concluded that exceeding a $3- \SI{5}{\degreeCelsius}$ temperature difference between two cells connected in parallel can lead to a reduced cycle lifetime \cite{rahn_Battery_2013}. Therefore, understanding the extent of internal heat activity within the cell is crucial for overcoming these challenges \cite{kim_Modelling_2011,yang_Realtime_2013,sethuraman_Realtime_2012}, as factors such as increased charge and discharge rates can significantly impact this internal heat generation and complicate effective thermal management \cite{lamb_MicroOptics_2020}. Good temperature management is therefore essential for optimal performance and cell longevity. Previous studies have often assumed constant temperature distributions on cell level, although this is typically not the case \cite{werner_Inhomogeneous_2020}. Variations in temperature distribution within the cell can lead to uneven reactions and localized aging, underscoring the importance of both accurate modeling and improved information about the actual thermal state of the cell.		
	Furthermore, it has been shown that not only the absolute temperature, but also temperature gradients within the cell can influence degradation processes, internal resistance and overall cell performance \cite{fleckenstein_Aging_2012, cloos_Challenges_2024, cloos_Thermal_2024}. For instance, significant temperature differences within the cell may accelerate aging in certain regions, leading to reduced capacity and efficiency over time. Given these critical temperature dependencies, the direct measurement of internal temperature provides valuable insights into the cell’s thermal behavior under operating conditions. This is particularly crucial for automotive applications, where cells are becoming larger and temperature gradients between the cell surface and core can impact performance and safety \cite{lobberding_Cell_2020}. In this context, newly introduced cell designs, such as CATL's cell-to-pack approach and BYD's blade battery, increase energy density by utilizing large-format cell structures \cite{byd_BYD_2022, catl_Innovative_2022}. However, due to their inherently larger cell format, these designs also intensify challenges in managing temperature gradients within the cell. This makes advanced temperature monitoring even more critical to ensure safe and efficient operation. To address this need, the integration of electronics and sensor technologies directly within the cell enables precise and unaltered measurements, extending battery monitoring to a new scale.
	
	These challenges underline the importance of advanced techniques that go beyond surface-level measurements to achieve a comprehensive understanding of internal cell temperatures. Such methods are essential to address the limitations of surface-based models and enable a more accurate prediction of thermal behavior within \acrshort{acr:LIB}s. In previous works, methods and models were investigated to derive the internal temperature of a cell from its surface temperature \cite{richardson_Battery_2014}. For this purpose, a point surface temperature and impedance were measured for each cell. In order to obtain sufficient information about the temperature distribution within the entire cell, a large number of sensors would be required. Moreover, the external measurement of temperature is not sufficient. The heat distribution on the surface of cells might not be homogeneous during operation and the external environment can influence the results \cite{veneri_Technologies_2017}. For example, the temperature at the cathode is generally higher than at the anode due to the lower electrical conductivity of the cathode material \cite{kim_Modelling_2011,spitthoff_Peltier_2021}. Therefore, even when using a thermal model, a point measurement of the surface temperature is not sufficient to reliably estimate the internal temperature distribution \cite{richardson_Battery_2014}. 
	
	Fiber optic measurement techniques offer decisive advantages in this context \cite{wahl_importance_2021}. They provide precise, spatially resolved data on temperature and pressure along the entire fiber, enabling a detailed understanding of thermal dynamics. Accurate knowledge of the temperature profile at the cell surface allows for better estimates of the temperature distribution inside the cell. Fiber optic sensors can be integrated with minimal impact due to their small diameter usually less than $\SI{200}{\um}$. This is particularly beneficial because measuring the temperature of clamped cells with conventional thermocouples can be challenging. The geometry of thermocouples often results in localized pressure maxima when pressed against the cell surface, which can induce defects and accelerate aging processes \cite{willenberg_development_2020}.
	Building on these practical advantages, the unique characteristics of optical fibers further enhance their potential as versatile sensors. Commonly used for data transmission, optical fibers rely on their optical conductivity to transmit signals via induced light. However, external disturbances, such as variations in temperature or pressure, can also alter the signal, providing valuable additional information about the external environment. With a suitable measurement technology, this behavior enables distributed sensing along the entire fiber. Glass fibers, in particular, are practically unaffected by electromagnetic fields \cite{han_review_2021}. This makes them an ideal choice for use as sensors in batteries, as they are not influenced by external electromagnetic interference. Furthermore, the chemical inertness of the glass fiber coating allows seamless integration into the chemically aggressive environments found within battery cells. 
	Because of these advantageous properties, various studies have explored the integration of fiber optic sensors into battery cells to measure internal states, such as changes in electrolyte mass composition, electrode expansion and temperature variations \cite{nedjalkov_Refractive_2019}. Similar to the approach presented in this work, \textit{Yu et al.} demonstrated distributed fiber optic measurements to capture internal in-plane temperature differences in lithium-ion cells for both pouch and cylindrical configurations \cite{yu_Distributed_2022, yu_Distributed_2021}. Latest reviews highlight the wide range of possibilities to use the sensors for the application in lithium ion batteries \cite{han_review_2021}. However, none of this work were able to sufficiently fulfill spatially resolved measurements of temperatures for use in lithium-ion battery cells.
	
	In this paper, we demonstrate the potential of fiber optic sensors for spatially resolved temperature measurement. Our methodology exploits general backscatter effects in glass fibers, enabling the detection of both thermal and mechanical influences. However, in the scope of this work, we will focus specifically on the thermal aspects. We investigate the spatial resolution capabilities of these fibers, assessing their performance across different scenarios. 
	Additionally, we examine the influence of adhesive points and sealing seams, which are essential steps for the integration of fiber optic sensors into batteries.

	\section{Fundamentals}
	\label{kap:Fundamentals}
	
	\subsection{Principle of Fiber Optic Sensors}
	\label{sec:FiberOpticSensors}
	
\begin{figure}[ht]
		\hspace*{\fill}%
		\subcaptionbox{\label{fig:Fiber}}{\includegraphics[width=0.45\textwidth]{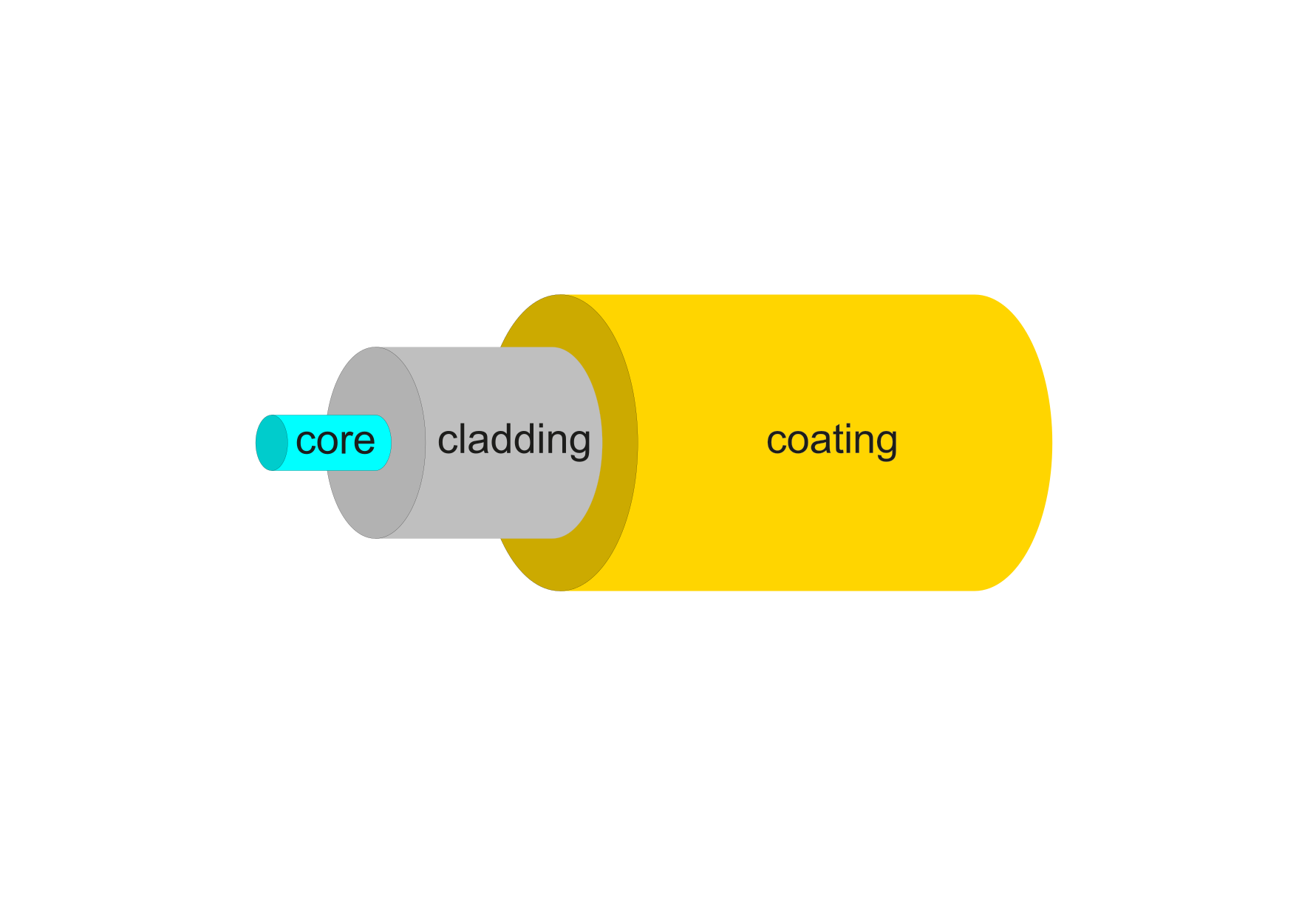}}\hfill%
		\subcaptionbox{\label{fig:FBG}}{\includegraphics[width=0.45\textwidth]{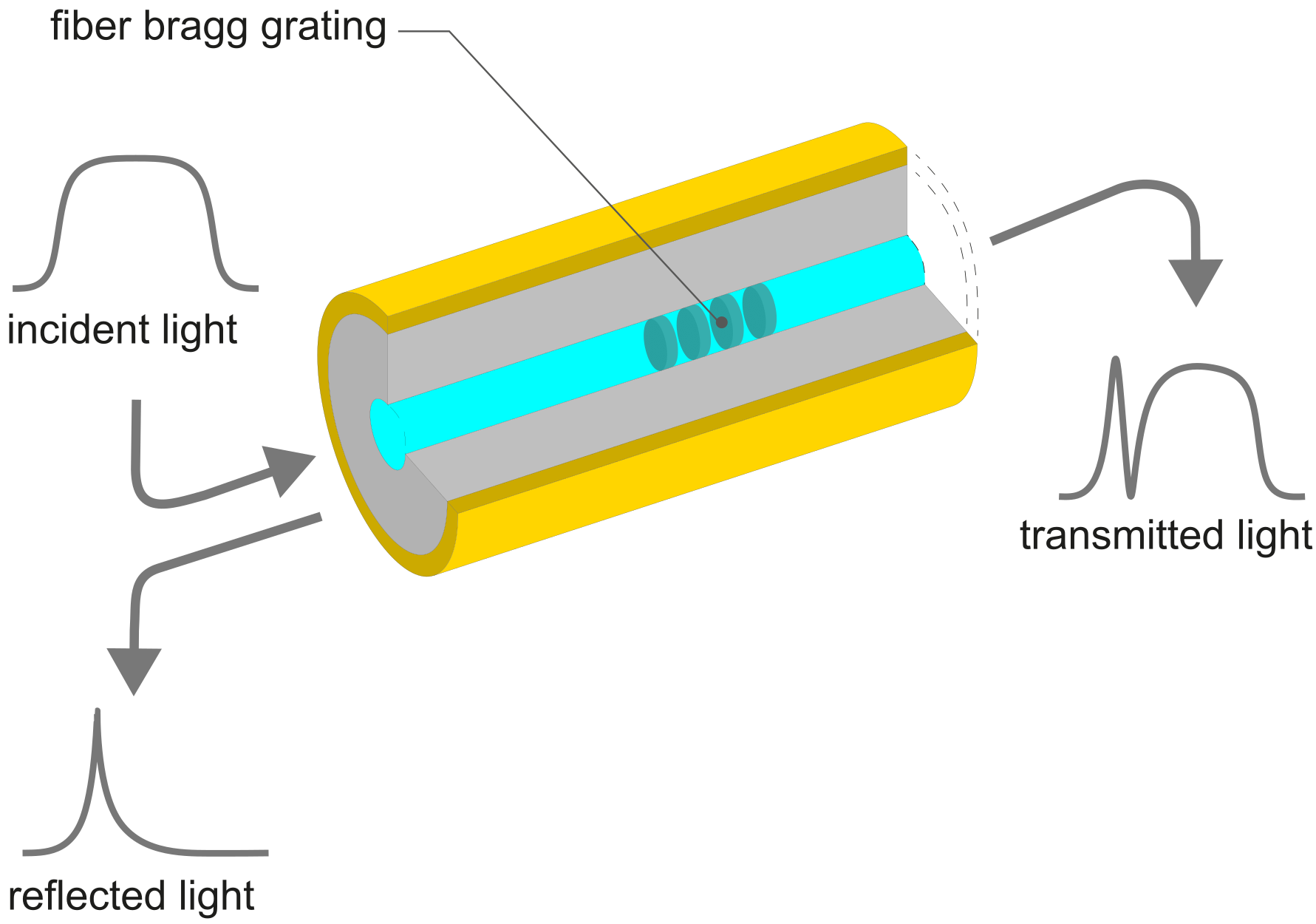}}%
		\hspace*{\fill}%
		
		\smallskip
		
		\hspace*{\fill}%
		\subcaptionbox{\label{fig:Beam}}{\includegraphics[width=0.45\textwidth]{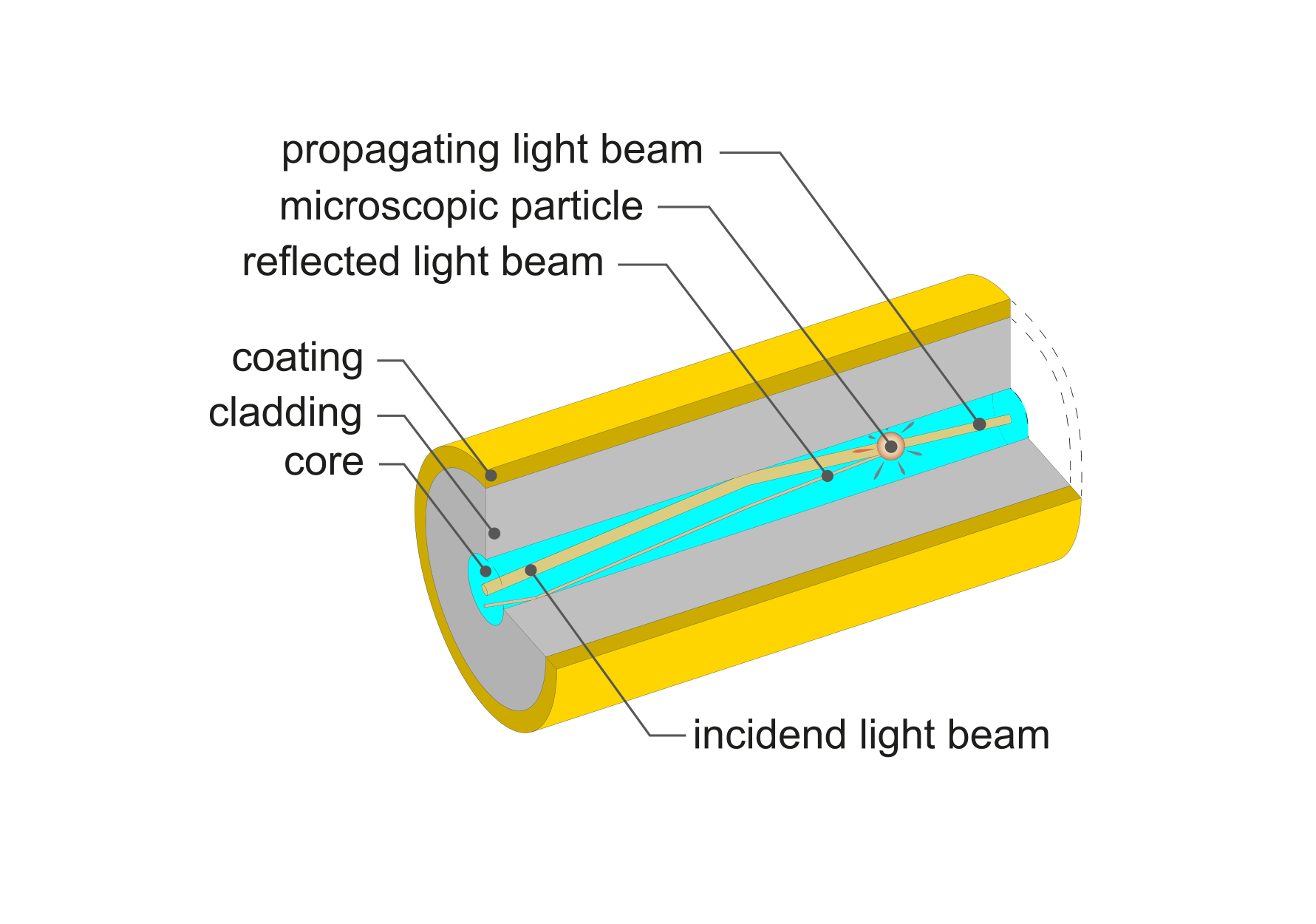}}\hfill%
		\subcaptionbox{\label{fig:Rayleigh}}{\includegraphics[width=0.45\textwidth]{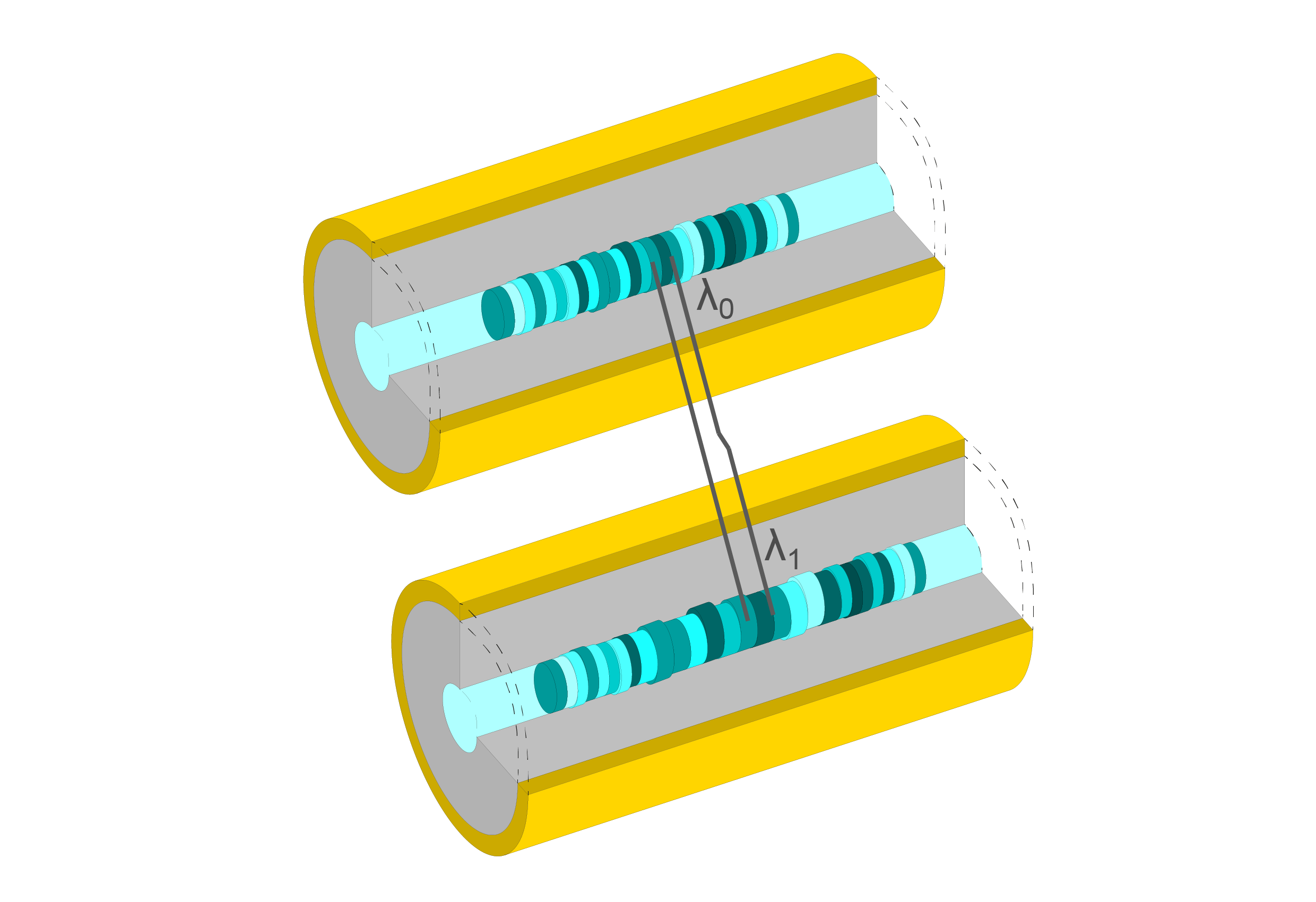}}%
		\hspace*{\fill}%
		\caption{Schematic representation of optical fibers and light propagation or refraction properties: (a) Structure of an optical fiber, showing the core, cladding and coating. (b) Reflection principle of an injected light signal into a fiber with an imprinted FBG. (c) Rayleigh backscatter effect from microscopic particles. (d) Division of the fiber into material-characteristic quasi-segments and corresponding refractive index variations without ($\mi{\lambda}{0}$) and with ($\mi{\lambda}{1}$) external influence on the fiber.}  
		\label{fig:Glasfaser}
	\end{figure}

	The sensitive properties and advantages of the material resistance of optical fibers have led to fiber optic sensors becoming increasingly established in the field of measurement sensor technology in recent years \cite{han_review_2021}. Essentially, a fiber is a relatively simple structure consisting of a core, cladding and coating, as illustrated in Figure~\ref{fig:Fiber}.

	A key feature of fiber optic sensors is the number and distribution of measuring points. These systems can be categorized into discrete, distributed and quasi-continuous measurement systems \cite{glisic_Fibre_2007}. For discrete measurement, there are a number of options that will not further be discussed in detail. Instead, the focus is shifted to the underlying principles of distributed and quasi-continuous sensors, which are mostly based on the superimposed comparison of two coherent light beams. These beams must have the same wavelength or a wavelength shifted by $2\pi$ to ensure maximum constructive interference between them. Additionally, they have traveled the same path through one or two identical optical fibers. If an external event such as strain or temperature change affects the optical fiber, this results in a phase shift of the imprinted wavelength. Interferometric measurement devices are capable of detecting a phase shift $\phi$ due to changes in refraction index $n$, wavelength $\lambda$ and path length $L$ \cite{peters_Fiber_2014,mikolajek_Temperature_2020}:
	\begin{equation}
		\Delta \phi = \frac{2 \pi}{\lambda} \cdot n \cdot L
		\label{eq:phaseshift}
	\end{equation}
	Common examples of interferometric systems include the Fabry-Pérot interferometer and the Michelson interferometer \cite{corke_Allfibre_1983, wu_high_2012, yuan_Fiber_2012, lee_Interferometric_2012}.
	
	In recent years, the advancement in optical measurement technology has led to the development of \acrfull{acr:FBG} distributed sensor technologies, offering an efficient and precise alternative to traditional measurement methods that utilize interferometers. These sensors have found applications in ultra-high temperature sensing in harsh environments, as well as in monitoring strain and temperature in battery applications \cite{azhari_novel_2014, fleming_Development_2018,peng_High_2019,novais_Internal_2016}. While interferometers rely on the phenomenon of light wave interference and often require complex setups, Fiber Bragg Sensors offer another solution for measuring various physical parameters. \acrshort{acr:FBG} are periodic optical gratings imprinted into fiber cores through complex techniques \cite{elsmann_Faserbragggitter_2017, fang_Fundamentals_2012}. These gratings create a modulation in the refractive index along the fiber, enabling selective reflection of light near the Bragg wavelength. When a light pulse with a broad frequency spectrum passes through the fiber, only rays close to the Bragg wavelength are reflected by the \acrshort{acr:FBG}, as illustrated in Figure~\ref{fig:FBG}. This selective reflection allows for precise temperature and strain measurements while keeping attenuation losses low \cite{mikolajek_Temperature_2020}.
	Previous studies have integrated \acrshort{acr:FBG} sensors in lithium-ion battery cells for applications such as internal temperature and strain monitoring \cite{novais_Internal_2016}, as well as electrolyte composition analysis \cite{nedjalkov_Refractive_2019,su_Fiber_2021}. For instance, \textit{Peng et al.} demonstrated high-precision strain measurements in pouch cells using \acrshort{acr:FBG} sensors \cite{peng_High_2019}. 
	
	Nevertheless, \acrshort{acr:FBG} sensors have a disadvantage compared to other techniques outlined: The width of the gratings is not arbitrarily adjustable and accordingly focuses only on individual sections along the fiber. It is thus not possible to measure along the entire fiber. This is when general backscattering effects become relevant.

	\subsection{Backscattering Effects}
	\label{sec:Backscatter}

	Some optical measurement methods rely on general backscattering effects, which enable quasi-continuous measurements along the entire length of the sensor fiber. These techniques fundamentally differ from Fiber Bragg Gratings (FBGs) in their physical mechanism. Backscatter-based systems derive their functionality from the intrinsic scattering phenomena within the fiber material, such as Rayleigh, Raman, or Brillouin scattering, rather than relying on periodic refractive index modulations as in FBGs. When light propagates through an optical fiber, imperfections and microscopic irregularities within the material scatter portions of the light back toward its source. This backscattered light inherently carries information about local changes in temperature and strain along the fiber. Despite the difference in physical origin, the functional principle of extracting spatially resolved temperature and strain data is conceptually similar to that of FBGs. The continuous nature of the backscattering measurement enables high spatial resolution over long distances, offering a complementary approach to the discrete sensing points provided by FBGs.
	
	The ability of backscattering systems to extract temperature and strain information stems from distinct physical interactions between light and the fiber. Rayleigh, Raman and Brillouin scattering are key mechanisms underlying these interactions, each contributing unique characteristics to the backscattered signal. These scattering processes occur randomly and in all directions, but sensing systems utilize only the backward-propagating fraction. While Raman and Brillouin scattering are an inelastic process, which utilize specific vibrational phenomena (molecular vibrations and acoustic phonons, respectively) to provide localized and highly sensitive environmental information, Rayleigh scattering operates through refractive index variations caused by structural imperfections and represents continuous signal propagation along the optical fiber. Latter is an elastic process where the incoming and outgoing wavelengths are identical, preserving photon energy but altering direction. This elasticity results in high scattering intensity, enhancing the signal-to-noise ratio in measurements \cite{peters_Fiber_2014,muanenda_Application_2019,kreger_High_2006, liehr_Fibre_2015}. Among these backscatter effects, Rayleigh scattering dominates light propagation losses in optical fibers due to interactions with particles much smaller than the light’s wavelength. Figure~\ref{fig:Beam} illustrates this mechanism, showcasing how light interacts with these particles to produce the backscattered signal. 

	Upon using a commercially available glass fiber sensor, variations in the intensity profile of Rayleigh scattering are observed along the fiber \cite{luna_Highdefinition_2022}. This profile remains highly stable when the measurement is repeated under unchanged external conditions, thus representing a distinctive pattern for a specific fiber segment. The stability is attributed to the nature of Rayleigh scattering, which arises from the elastic scattering process at local defects, distortions in waveguide geometry distortions or refractive index variations, for instance, caused by the solidification of the glass during the production process, see Figure~\ref{fig:Rayleigh} \cite{hartog_Introduction_2017}. This results in a division of the fiber into quasi-segments in which the scattering varies from one segment to another but remains stable under constant conditions \cite{soller_High_2005}. Introducing changes in temperature or strain of the fiber leads to spatial stretching or compression of the pattern. This phenomenon forms the basis of Rayleigh sensor technology, wherein alterations in the local Rayleigh pattern can be converted into local temperature or strain values by analyzing the wavelength shift of $\mi{\lambda}{0}$ to $\mi{\lambda}{1}$ according Figure~\ref{fig:Rayleigh} \cite{beckers_Basics_2017,samiec_Verteilte_2011}. However, this is similar to Fiber Bragg gratings, yet it provides the additional benefit of not necessitating expensive additional fiber preparation and the measurement is not spatially limited. This is due to the quasi-segments along the glass fiber mimicking a continuous grating structure, containing information along the entire length of the fiber.
	
	For an understanding of the measurement system, a brief overview of the methodology will be presented below.

	\subsubsection{Optical Time Domain Reflectometry (OTDR)}
	\label{sec:OTDR}
	
	\acrfull{acr:OTDR} is a fiber optic measurement technique based on backscattering effects. In this method, a short, high-power light pulse is transmitted through the optical fiber. The backscattering is then detected at the fiber's entry point using a photodiode coupled with a circulator or splitter and the time taken for the signal to return is recorded. The resulting waveform is digitized and subjected to analysis using a data acquisition system. To achieve precise calculations of the propagation time of backscattered signals and accurately locate backscattering events along the fiber, it is synchronized with the optical pulse source.
	The position $x$ within the fiber corresponding to the backscattering event is determined using the following formula \cite{lu_Distributed_2019}:
	
	\begin{equation}
		x = \frac{v}{2n} \cdot t
		\label{eq:streuposition}
	\end{equation}
	
	The time $t$ taken for the backscattered signal with the refractive index $n$ of the glass fiber to be detected after the pulse is transmitted. During this process, the light travels twice the distance, with respective velocities of light $v$ in the fiber. This \acrshort{acr:OTDR} technique is widely used for gaining valuable insights into the characteristics and integrity of optical fibers, making it a crucial tool for various fiber optic applications. The pulse duration or width is a critical characteristic of \acrshort{acr:OTDR} and significantly influences the measurement resolution \cite{lu_Distributed_2019}.
	The performance of \acrshort{acr:OTDR} is evaluated based on three key parameters: the dynamic range (maximum detectable fiber loss), spatial resolution (location accuracy of resolving adjacent events) and dead-zone \cite{bao_Recent_2017}. The extent of the dead-zone is influenced by factors like the width of the incoming optical pulses and the bandwidth of the photodetector \cite{mao_Sensing_2020}. To overcome this limitation, alternative methods such as \acrfull{acr:OFDR} have been introduced.

	\subsubsection{Optical Frequency Domain Reflectometry (OFDR)}
	\label{sec:OFDR}
	
	In contrast to the \acrshort{acr:OTDR} method, \acrfull{acr:OFDR} employs a tunable laser as the light source, rather than using short light pulses. The tunable laser emits highly coherent light with a broad wavelength range. With the help of a linear optical frequency sweep, the entire length of the glass fiber can be used as a sensor, making it a valuable tool for various fiber optic applications. The OFDR setup involves splitting the laser beam into a test arm, directed towards the optical fiber sensor and a reference arm, which form a Mach-Zehnder interferometer \cite{lee_Interferometric_2012,yuksel_Analysis_2009}. 
	
	Backscattered signals from the test arm are then combined with the reference arm signal, leading to the formation of interference patterns based on the phase differences of the backscattered light. The following used \acrshort{acr:OFDR} method relies on Rayleigh scattering, as previously described. For the temperature range relevant in this work, the refractive index of the fiber sensors can be assumed to be constant for a given wavelength of the incident light in the C-band ($ 1530 - \SI{1565}{\nm}$) \cite{tan_Temperature_2000}. Induced length changes (e.g. due to punctual increased temperature) in the fiber affect the backscattered light's travel time, which is detected with an optical detector. 
	Typically, the reference arm signal is much stronger, necessitating the use of balanced photodetectors to equalize the power difference \cite{kreger_High_2006}. The combined signal at the detector is then subjected to a Fast Fourier Transformation (FFT), breaking it down into its individual wavelength components. The wavelength difference corresponds to the reflection position, while the amplitude represents the intensity of the backscattering \cite{soller_High_2005}. Although this makes the measurement system quite complex, it significantly enhances its sensitivity.
	
	The temperature change $\Delta T$ and/or strain $\epsilon$ acting on the fiber and its effect on the reflected wavelength can be described by the following formula \cite{mikolajek_Temperature_2020,kreger_High_2006}: 
	
	\begin{equation}
		\begin{array}{ll}
			\frac{\Delta\lambda}{\mi{\lambda}{0}} = \mi{K}{\epsilon} \cdot \epsilon + \mi{K}{T} \cdot \Delta T & , \ \mi{K}{T} = \mi{\alpha}{\Lambda} + \mi{\alpha}{n}
		\end{array}
		\label{eq:Wellenlängenabhängigkeit}
	\end{equation}
	
	$\Delta \lambda$ describes the wavelength change $\frac{\si{\um}}{\si{\m}}$ in $\mu\epsilon$, $\mi{\lambda}{0}$ the initial irradiated wavelength, $\mi{K}{e}$ the material-specific strain dependency and $\mi{K}{T}$ the temperature coefficient. The latter includes both the thermo-optical coefficient $\mi{\alpha}{n}$, i.e. the change of the material-specific refractive index due to the temperature change and the thermal-expansion coefficient $\mi{\alpha}{\Lambda}$, which describes the temperature dependence due to the expansion of the fiber \cite{mikolajek_Temperature_2020,sirkis_Unified_1993,odwyer_Thermal_2004}. As shown in Formula~\ref{eq:Wellenlängenabhängigkeit}, strain and temperature both influence the wavelength, which complicates distinguishing between the two effects. This intrinsic weakness is shared by most fiber optic measurement methods that record the length change of the sensor during the measurement. 
	
	With known environmental conditions, a calibration curve and thus reference values for $\mi{K}{\epsilon}$ and $\mi{K}{T}$ can be determined. As a result, using Formula~\ref{eq:Wellenlängenabhängigkeit}, the external influence on the fiber can be inferred. With the \acrshort{acr:OFDR} method, the wavelength change can be measured, enabling the determination of the location and magnitude of the sensor's change. However, this measurement method does not provide conclusive information about the specific physical nature of the induced load, such as pressure or tension. Due to the proposed high spatial resolution required when using fiber optic sensors for state monitoring in and on batteries, the \acrshort{acr:OFDR} measurement method, utilizing Rayleigh backscattering, is well-suited.
	
	To validate this, we conduct a series of tests focusing on the capability of various single-mode glass fibers to reliably detect temperature changes while ensuring spatial resolution sufficient to identify thermal anomalies. Furthermore, our study provides preliminary findings on the application of these sensors in \acrshort{acr:LIB} cells. These results could pave the way for using fiber optic sensors in battery state monitoring and predictive diagnostics.
		
	\newpage

	\section{Experimental Set-Ups and Thermal-Characterization of Fiber Sensors}
	\label{kap:Set-Up}
	
	To investigate the influence of temperature on fiber optic sensors, we conducted a series of experiments using various setups under defined conditions to comprehensively analyze and describe their thermal behavior. Additionally, the influence of specific steps of fiber integration in \acrshort{acr:LIB}s on the functionality and performance of these fiber optic sensors is analyzed.
	
	In the following, a Rayleigh \acrshort{acr:OFDR} measurement technique, as previously described, using an \textit{ODiSI 6104} instrument was performed \cite{luna_Highdefinition_2022}. The details of the device are outlined in its data sheet, as summarized in Table~\ref{tab:MeasurementAccuracy} \cite{luna_Highdefinition_2022}. The device is capable to measure a wide range of fiber types, which will be further specified. The \textit{ODiSI 6104} has the ability to measure fiber optic sensors on four different channels, which allows us to measure a maximum of four fiber optic sensors at the same time. The maximum sensor length of the measurement device is $\SI{20}{\m}$ and can be possibly increased up to $\SI{100}{\m}$ with an extended range remote module. Moreover, it offers a maximum spatial resolution of $\SI{0.65}{\mm}$ and sampling rate of $\SI{250}{\hertz}$ \cite{luna_Highdefinition_2022}. 
	
	The device does not directly convert the measured changes into temperatures; therefore, only the relative change in length of the fiber under test is considered. The ratio between the original elongation of the glass fiber and the elongation caused by the mechanical and thermal load is referred to as strain. Although strain generally describes a dimensionless quantity, $\epsilon$ is usually used as the unit, also described in the previous chapter~\ref{sec:OFDR}. In the subsequent sections of this work, we will further elaborate on the correlation between temperature and the length change of the used fibers.
	
	\begin{table}[ht]
	\caption{Measurement accuracy and resolution of \textit{ODiSI 6104} with set spatial resolution of $\SI{2,6}{\mm}$ specified by the manufacturer \cite{luna_Highdefinition_2022}.}
	\footnotesize
	\centering
	\begin{tabular}{ll}
		\toprule
		Parameter & Value in \si{\mu\epsilon} \\ 
		\midrule
		Resolution & $0.1$ \\
		Instrument accuracy & $1$ \\
		System (instrument and sensor) accuracy & $\pm 30$ \\
		Measurement uncertainty at zero strain & $\pm 2$ \\
		Measurement uncertainty across full strain range & $\pm 2$ \\
		Measurement uncertainty & $\pm 0.6$ \\ 
		\bottomrule
	\end{tabular}
	\label{tab:MeasurementAccuracy}
	\end{table}

	For thermal characterization of our fibers, we used a \textit{LabEvent L T/64/40/3} temperature chamber from \textit{Weiss Technik GmbH}, as detailed in Table~\ref{tab:Temperaturechamber} \cite{weisstechnik_Technical_2021}. This chamber regulates the set temperature using an integrated sensor with a maximum deviation of $\SI{\pm 2}{\degreeCelsius}$.
	
	\begin{table}[ht]
	\caption{Technical details of used temperature chamber \textit{LabEvent L T/64/40/3} of \textit{Weiss Technik GmbH} \cite{weisstechnik_Technical_2021}.}
	\footnotesize
	\centering
	\begin{tabular}{ll}
		\toprule
		Parameter & Value in \si{\degreeCelsius} \\ 
		\midrule
		Temperature range & $-40$ to $180$ \\
		Temperature deviation, temporal & $\pm 0.3$ to $\pm 1$ \\
		Temperature deviation, spatial & $0.5$ to $2$ \\
		\bottomrule
	\end{tabular}
	\label{tab:Temperaturechamber}
	\end{table}
	
	Two different temperature sensors were used to record reference temperature values. We used a calibrated \textit{PT1000} with a measurement range of $\SI{-50}{\degreeCelsius}$ to $\SI{280}{\degreeCelsius}$ and an accuracy of $\SI{\pm 2}{\degreeCelsius}$ \cite{industries_Platinum_2023}. Additionally, verification was conducted using a calibrated \textit{AD590} from \textit{Analog Devices}. This sensor has a temperature range of $\SI{-55}{\degreeCelsius}$ to $\SI{150}{\degreeCelsius}$ with a resolution of $\SI{\pm 0.5}{\degreeCelsius}$ and a maximum deviation of $\SI{\pm 0.8}{\degreeCelsius}$ \cite{analogdevicesinc._Data_2013}.
		
	When selecting suitable singlemode glass fibers for the fabrication of fiber optic sensors, one can chose from a wide range of fibers featuring a wavelength in C-band. In the following section, seven types of glass fibers are investigated and characterized based on their temperature sensitivity. All used fibers have a core diameter of $\SI{9}{\um}$ and a cladding diameter of $\SI{125}{\um}$, the total diameter including the coating, however, differs between around $\SI{155}{\um}$ to $\SI{200}{\um}$. To manufacture sensors from these fibers, they must first be prepared with a connector for the measuring device and a termination at the other end. The termination is achieved by splicing a coreless, acrylic-coated fiber (approximately \SI{2}{\cm} long), which prevents reflected light from interfering with measurements and causing backscatter effects. On the device side, an LC/APC connector is used, typically available as a commercially sold pigtail and is also attached via a splicing process. The device used for this is the \textit{OFS-95} from \textit{ShinewayTech} \cite{shinewaytech_OFS95S_2020}. In addition to preparing and manufacturing our own sensors, a comparison with commercially available \textit{High-Defintion Fiper Optic Strain Sensors} from \textit{Luna Innovations}, hereinafter referred to as Fiber A, was also carried out \cite{luna_Highdefinition_2022}. Besides the thickness or type of coating, the composition of the glass can also have an impact. We therefore tested six different fibers, which are listed with their properties in Table~\ref{tab:listofsensors}. 
	
	\begin{table}[ht]
	\caption{List of examined optical fibers with a core thickness of $\SI{9}{\um}$, a cladding thickness of $\SI{125}{\um}$ \cite{luna_Highdefinition_2022,fibercore_Fibercore_2022}.}
	\footnotesize
	\centering
	\resizebox{\textwidth}{!}{%
		\begin{tabular}{lllll}
			\toprule
			ID & Material & Coating thickness in $\si{\um}$ & Temperature range in $\si{\celsius}$ & Product Specification \\ 
			\midrule
			Fiber A & Polyimide coating & 155 & -40 to +220 & Luna HD6S\\
			Fiber B & Polyimide coating & 155 & -40 to +220 & n.a.\\ 
			Fiber C & Dual acrylate coating \& Ge-doped core & 245 ± 7 & -55 to +85 & SM1500(9/125) \\ 
			Fiber D & Polyimide coating \& Ge-doped core & 155 ± 5 & -55 to +300 & SM1500(9/125)P \\
			Fiber E & Dual acrylate coating \& pure silica core & 245 ± 15 & -55 to +85 & SM1500SC(9/125) \\ 
			Fiber F & Polyimide coating \& pure silica core & 155 ± 5 & -55 to +300 & SM1500SC(9/125)P \\ 
			\bottomrule
		\end{tabular}%
	}
	\label{tab:listofsensors}
	\end{table}
	
	Fiber A is a high-definition optical fiber sensor with polyimide coating, specifically designed for high-precision strain and temperature measurements. It offers a wide temperature range of -40 to +220°C, making it suitable for a variety of applications in harsh environments. Fiber B has identical composition and properties to Fiber A, however, its length can be customized. Fiber C, D, E and F differ in their materials and composition and therefore also in their thickness and temperature range. Optical sensors were fabricated from all these fibers. 
	
	When connecting a sensor to the \textit{ODiSI 6104} for the first time, it is necessary to create a so-called custom key. This stores an individual Rayleigh backscatter information for the respective sensor. When generating the key, care must be taken that the thermal as well as mechanical conditions are the same. It is necessary to establish a well-defined reference point for each measurement by taring it with known environmental influences and interpreting the relative variations of the measurement signal with respect to this point. In this case, the fibers were tared in a non-stressed state at $\SI{20}{\degreeCelsius}$.
	
	As mentioned earlier, different spatial resolutions are possible. With decreasing spatial resolution, a higher sampling rate can be achieved. When observing measurements at rest, we noticed a high standard deviation $\sigma$. Subsequently, we conducted a comparison of different spatial resolutions ($\SI{0,65}{\mm}$, $\SI{1,3}{\mm}$, $\SI{2,6}{\mm}$) at a frequency of $\SI{1}{\hertz}$ using Fiber A as an example. External influences such as temperature and strain were kept homogeneous throughout the measurements. To evaluate the potential of achieving high spatial resolution (e.g., $\SI{2.6}{\mm}$) for thermal events using optical fibers, three independent cooling events were performed. A \textit{FREEZE 75} cooling spray from \textit{Kontakt Chemie}, capable of reaching temperatures as low as $\SI{-45}{\degreeCelsius}$, was used. The ambient temperature during the measurement was set to $\SI{25}{\degreeCelsius}$. For the experiments, a Fiber C was employed to capture the temperature changes with spatial precision. A zero-point calibration was performed at $\SI{25}{\degreeCelsius}$.
		
	To rule out the possibility that relaxation effects might influence the measurements, the thermal relaxation behavior of the Fiber C was assessed in a separate experiment. For this purpose, the fiber was placed in a temperature-controlled environment, as described previously. The climate chamber was set to $\SI{40}{\degreeCelsius}$ and allowed to homogenize before the fiber was introduced. After the chamber reached thermal equilibrium, a zero-point calibration was performed. The experiment was conducted over a duration of $\SI{48}{\hour}$, with continuous logging of sensor data.
	
\begin{figure}[ht]
	\centering
	\includegraphics[width=0.9\textwidth]{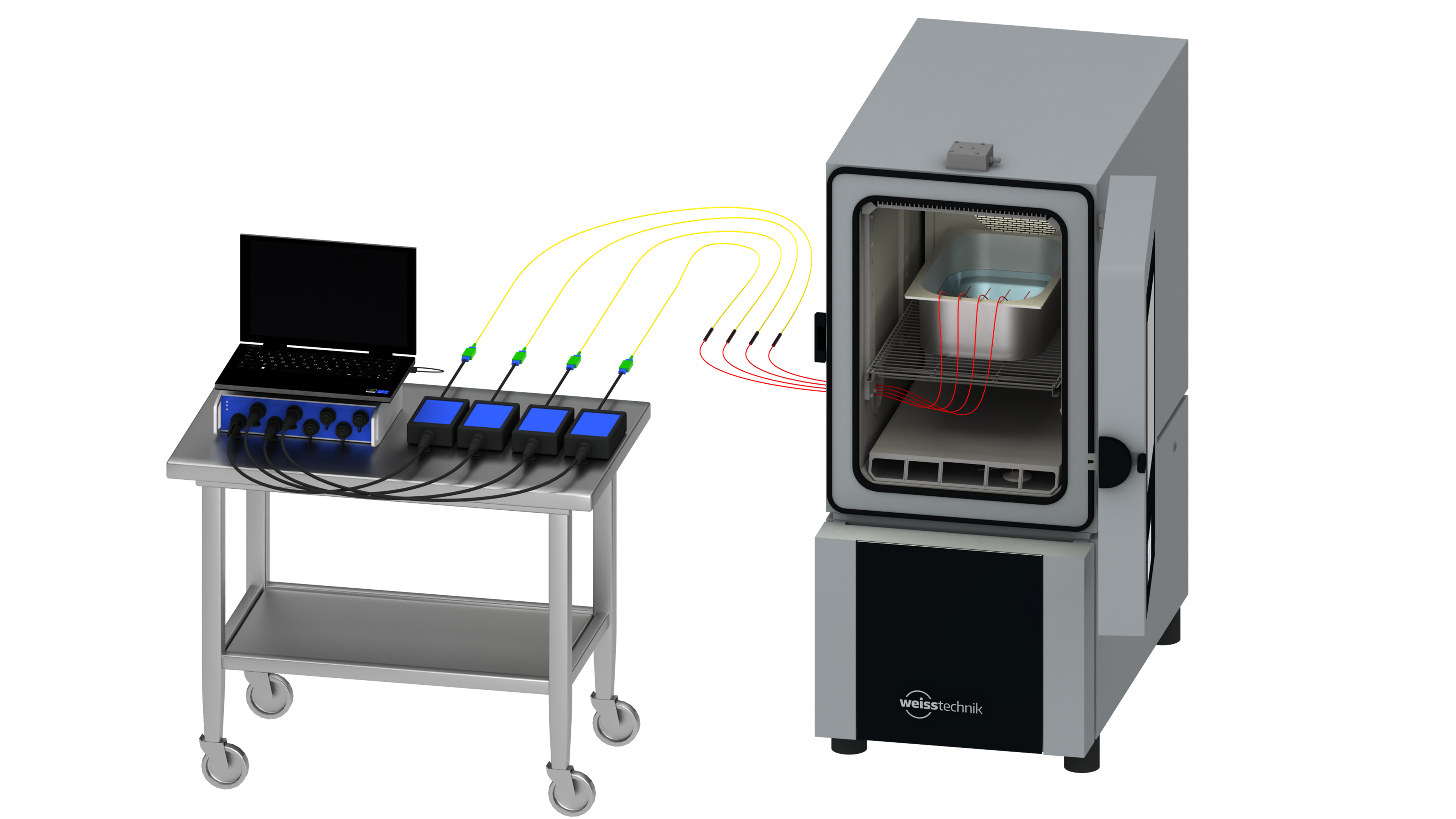}
	\caption{Exemplary test setups with a \textit{Weiss} climate chamber, a water basin for temperature measurement and an \textit{ODiSI 6104} interrogator for various experiments to determine the thermal dependency of different fibers.} 		
	\label{fig:MessaufbauBecken}
\end{figure}

	Due to the high sensitivity of our fiber sensors, further experiments were conducted to determine the temperature sensitivity of the fibers across a defined range of temperatures. It was observed that the measurement signal could be influenced by the active ventilation of the climate chamber. To eliminate this effect and ensure consistent characterization, the sensors were placed in a water container, which acted as both a thermal buffer and stabilizing medium, as shown in Figure~\ref{fig:MessaufbauBecken}.

	A stepwise temperature profile was applied, gradually increasing the temperature in $\SI{10}{\degreeCelsius}$ (or $\SI{5}{\degreeCelsius}$) increments from $\SI{5}{\degreeCelsius}$ to $\SI{80}{\degreeCelsius}$. At each step, the temperature was held constant for approximately $\SI{6}{\hour}$ to allow for the establishment of thermal equilibrium. A zero-point calibration was performed at $\SI{5}{\degreeCelsius}$ to ensure accurate baseline alignment for subsequent measurements. The water temperature was continuously monitored using reference sensors (\textit{PT1000} \& \textit{AD590}), providing a precise baseline for comparison with the fiber optic sensor measurements.  
	Since the fiber optic sensors are intended for use in or on batteries, the expected operating temperature range is limited by the thermal conditions typically encountered in battery systems. At cell temperatures above $\SI{90}{\degreeCelsius}$, critical conditions can arise, such as exothermic decomposition of electrolytes, which may eventually lead to safety-critical effects \cite{bandhauer_Critical_2011}. Temperatures below $\SI{5}{\degreeCelsius}$ are generally avoided when testing \acrshort{acr:LIB}, as they increase the viscosity of the electrolyte, reduce ion mobility and impair electrochemical efficiency, leading to performance loss and potential damage, a phenomenon also well described by physical models \cite{bihn_PhysicsBased_2024,singer_Kinetic_2017}. Therefore, the operational temperature range for the experiments was set between $\SI{0}{\degreeCelsius}$ or $\SI{5}{\degreeCelsius}$, depending on the specific test conditions and $\SI{80}{\degreeCelsius}$, to ensure alignment with typical battery system environments.
		
	To ensure accurate measurements and minimize potential interference from external factors, it is important to control other conditions that may impact the fiber optic sensor readings. One such dominant effect is the bending of optical fibers, which can significantly alter the measurement results. Increased bending can cause light to be gradually extracted from the fiber, influencing signal transmission and reducing measurement precision. As bending increases, it also affects the fiber’s numerical aperture, which in turn changes the light propagation within the fiber. Beyond a certain threshold, excessive bending can render the fiber’s signal transmission ineffective, making accurate measurements difficult or even impossible. Therefore, it is essential to carefully account for and minimize the effects of fiber bending during experiments.
	
	\begin{figure}[h]
		\hspace*{\fill}%
		\subcaptionbox{\label{fig:Alumodell}}{\includegraphics[width=0.30\textwidth]{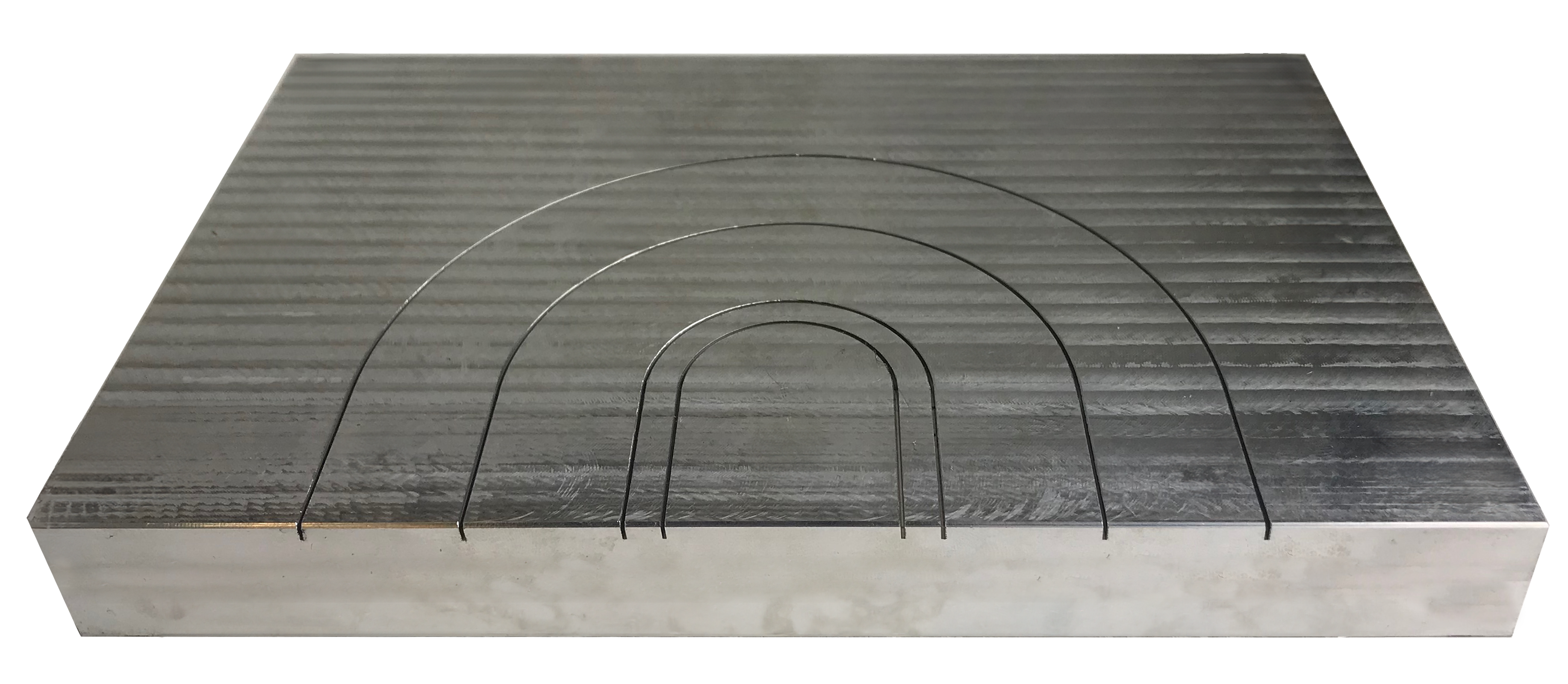}}%
		\subcaptionbox{\label{fig:MessaufbauAlu}}{\includegraphics[width=0.60\textwidth]{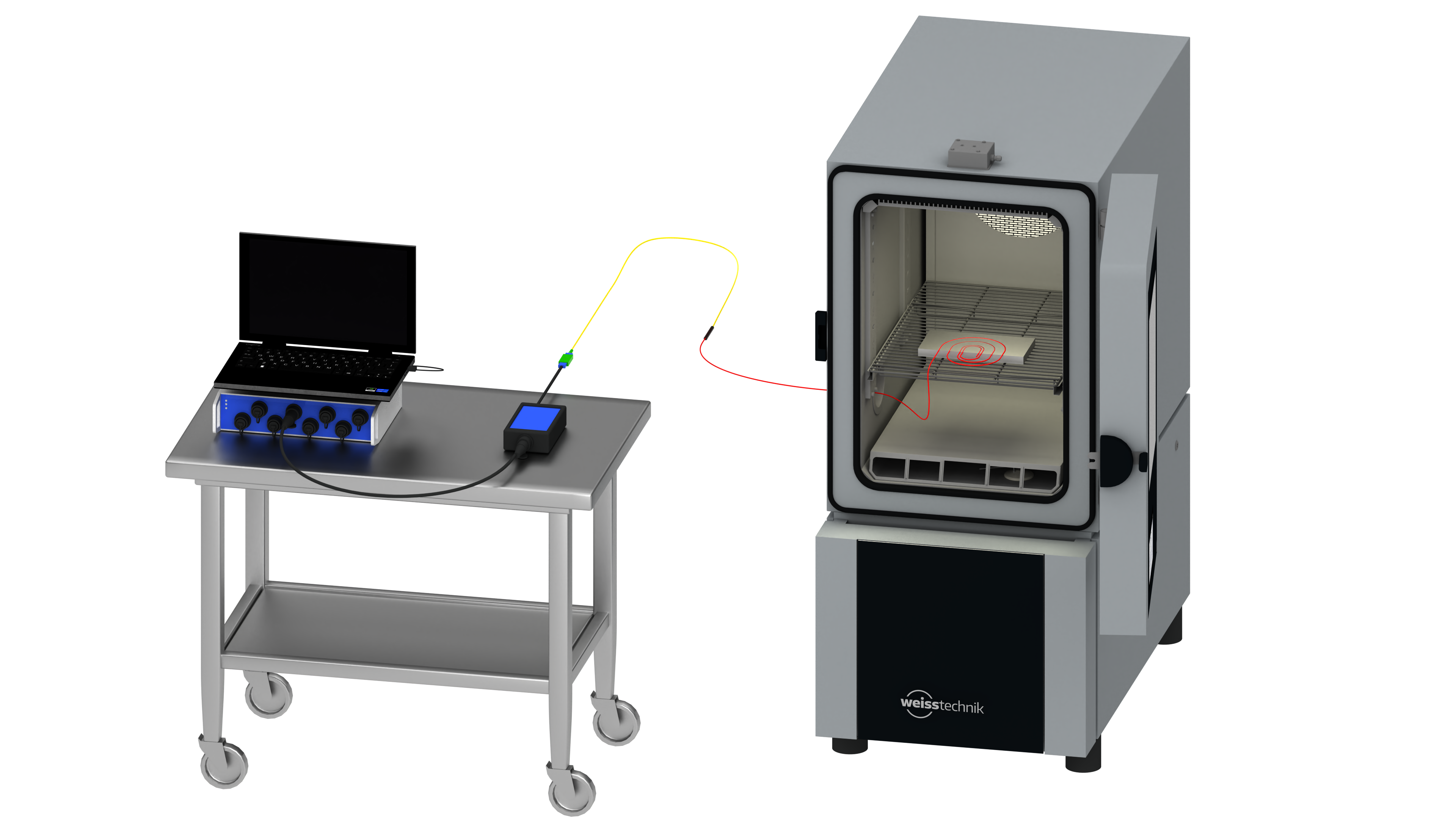}}%
		\caption{(a) Aluminum model with milled radii ($15, 20, 40$ and $\SI{60}{\mm}$) for measuring thermal dependence of fibers in bends. (b) Exemplary test setup with a \textit{Weiss} climate chamber, an aluminum plate with four milled bending radii and an \textit{ODiSI 6104} interrogator to determine the thermal dependence of bent fibers.}
	\end{figure}
	
	 In order to analyse these effects, several tests were conducted in the range of possible bending radii. According to \textit{Luna Innovations}, minimal bending radii up to $\SI{10}{\mm}$ are possible without distorting the measurement. However, tests have shown that a fiber fracture occurred at a radius of approximately $\SI{1}{\mm}$. It should be noted that although a fracture occurred only at a radius of $\SI{1}{\mm}$, micro-cracks in the fiber occur when the minimum bending radius specified by the manufacturer is exceeded \cite{chen_Crack_2017}. Therefore, we have allowed for a safety margin for future installations and testing and set the minimum radius to $\SI{15}{\mm}$. To test the fiber in a homogeneous bend, a model (Figure~\ref{fig:Alumodell}) was designed and milled into an aluminum block. Aluminum has a thermal conductivity of $235 \frac{W}{k \cdot m}$ and thus distributes heat significantly better than other possible materials such as conventional 3D printing materials such as \textit{PPA} or \textit{PA} (thermal conductivity of $< 1 \frac{W}{m \cdot k}$) \cite{ostermann_Anwendungstechnologie_2014,moldflowplasticslabs_Material_2007,eosgmbh-electroopticalsystems_PA2200_2024}. Latter material showed also poor temperature distribution and inhomogeneous measurement results in experiments. However, a significantly longer time constant for complete temperature acceptance was taken into account during the tests, since the size of the aluminum model also represents a heat reservoir.

	A milling depth and width of $\SI{0.5}{\mm}$ were selected for creating the grooves, ensuring that the fiber would not be crushed during installation while remaining almost entirely encased by aluminum to achieve optimal thermal contact. Radii of $15$, $20$, $40$ and $\SI{60}{\mm}$ were incorporated into the model. Unlike the previous setup, this experiment did not employ a water basin; instead, only the aluminum model was placed inside the temperature chamber, as shown in Figure~\ref{fig:MessaufbauAlu}.
	
	A Fiber C was threaded through all the grooves, allowing simultaneous monitoring of all radii in a single measurement. This setup ensured that the chamber's fan did not influence the results, as the fiber was shielded by the aluminum. The temperature steps ranged from $5$ to $\SI{80}{\degreeCelsius}$ with a zero-point calibration at $\SI{5}{\degreeCelsius}$, consistent with the earlier experiments. All the previously described experiments were conducted using a sampling rate of $\SI{1}{\hertz}$.

	\clearpage

	\section{Temperature Sensitivity of Fiber Sensors}
	\label{kap:TemperatureMeasurements}

	\textbf{Spatial Resolution and Measurement Precision}
	\newline
		
\begin{figure}[!b]
		\hspace*{\fill}%
		\subcaptionbox{\label{fig:ruhemessung065}}{\includegraphics[width=0.45\textwidth]{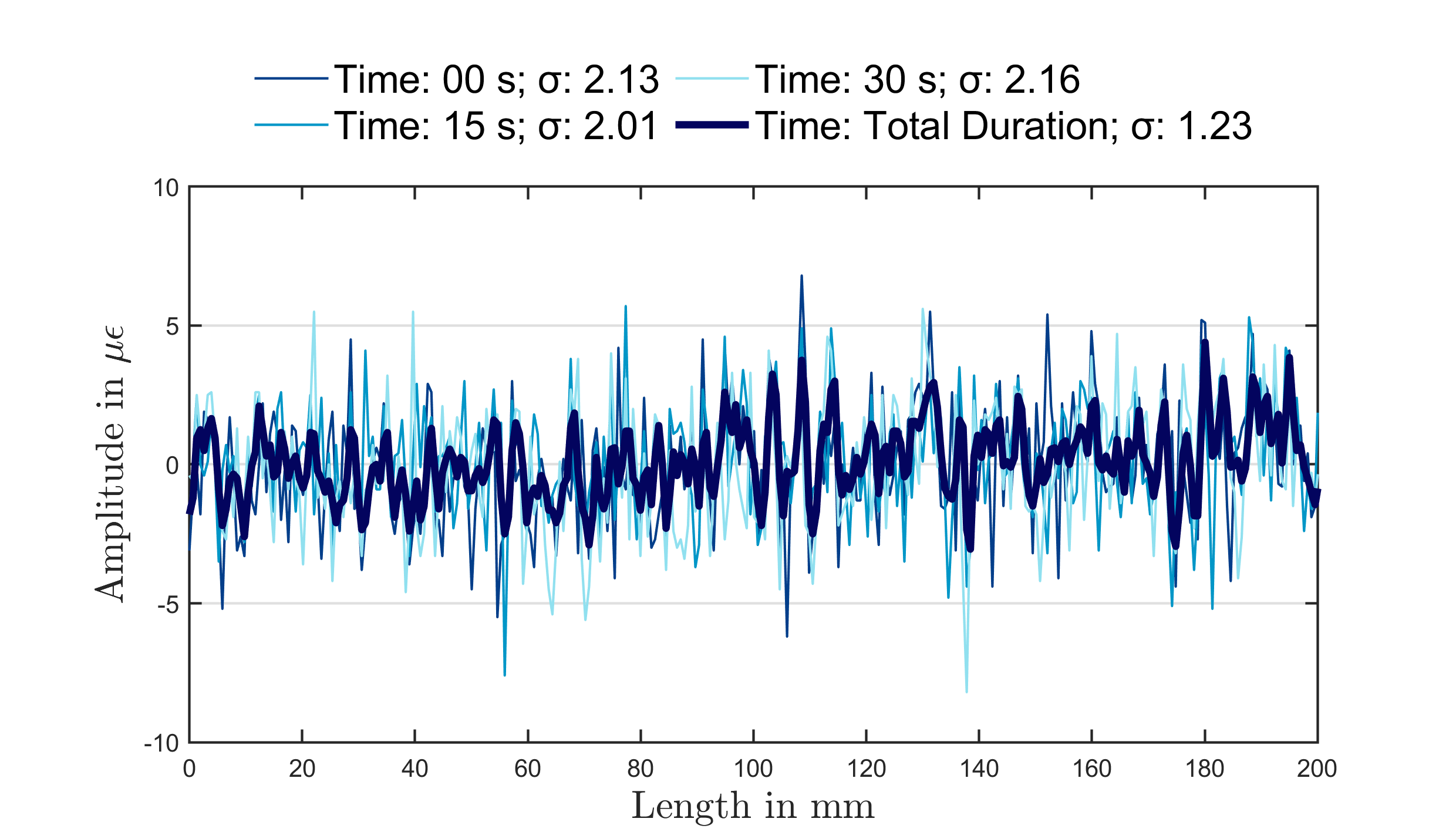}}\hfill%
		\subcaptionbox{\label{fig:ruhemessung13}}{\includegraphics[width=0.45\textwidth]{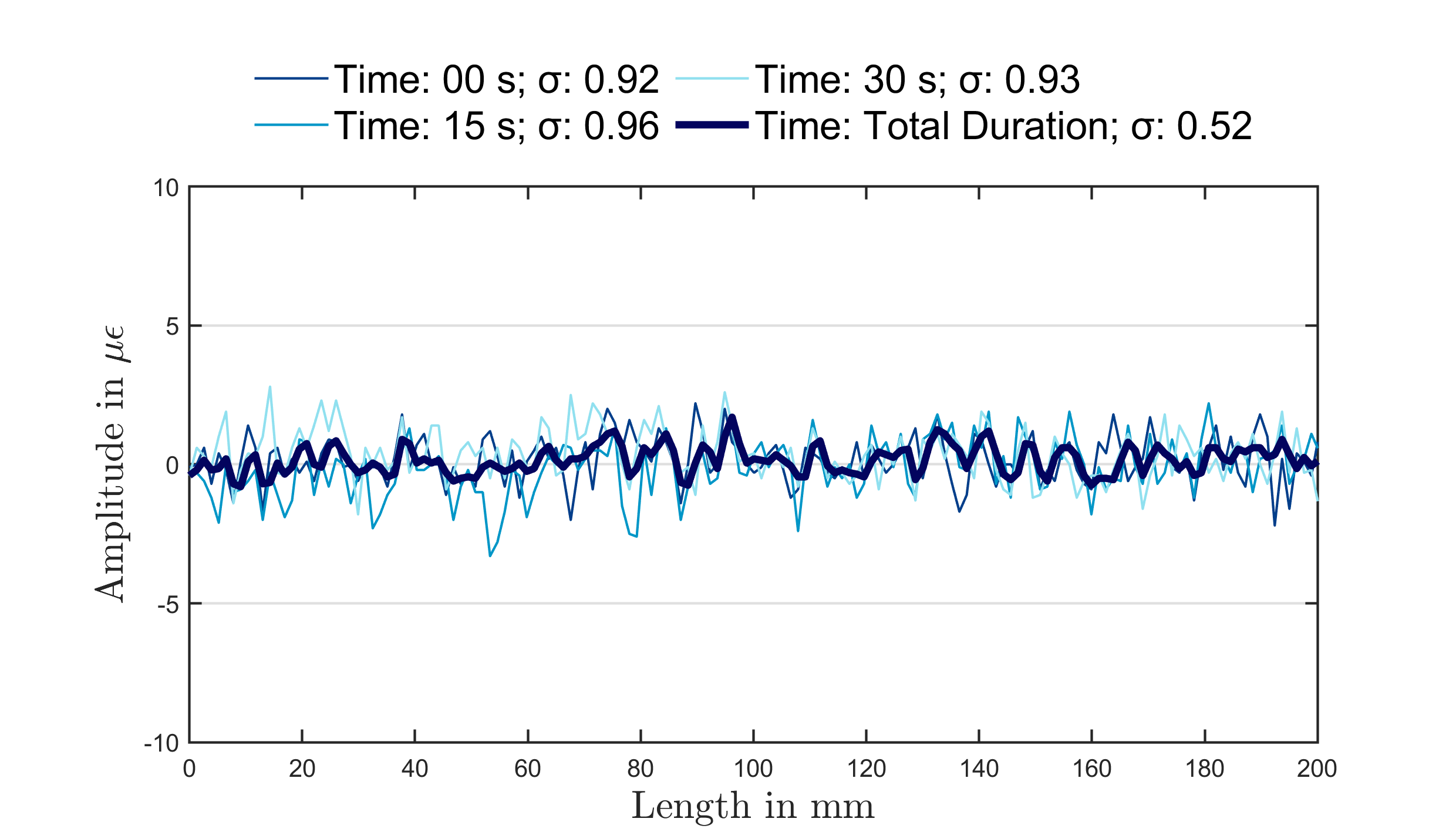}}%
		\hspace*{\fill}%
		
		\smallskip
		
		\hspace*{\fill}%
		\subcaptionbox{\label{fig:ruhemessung26}}{\includegraphics[width=0.45\textwidth]{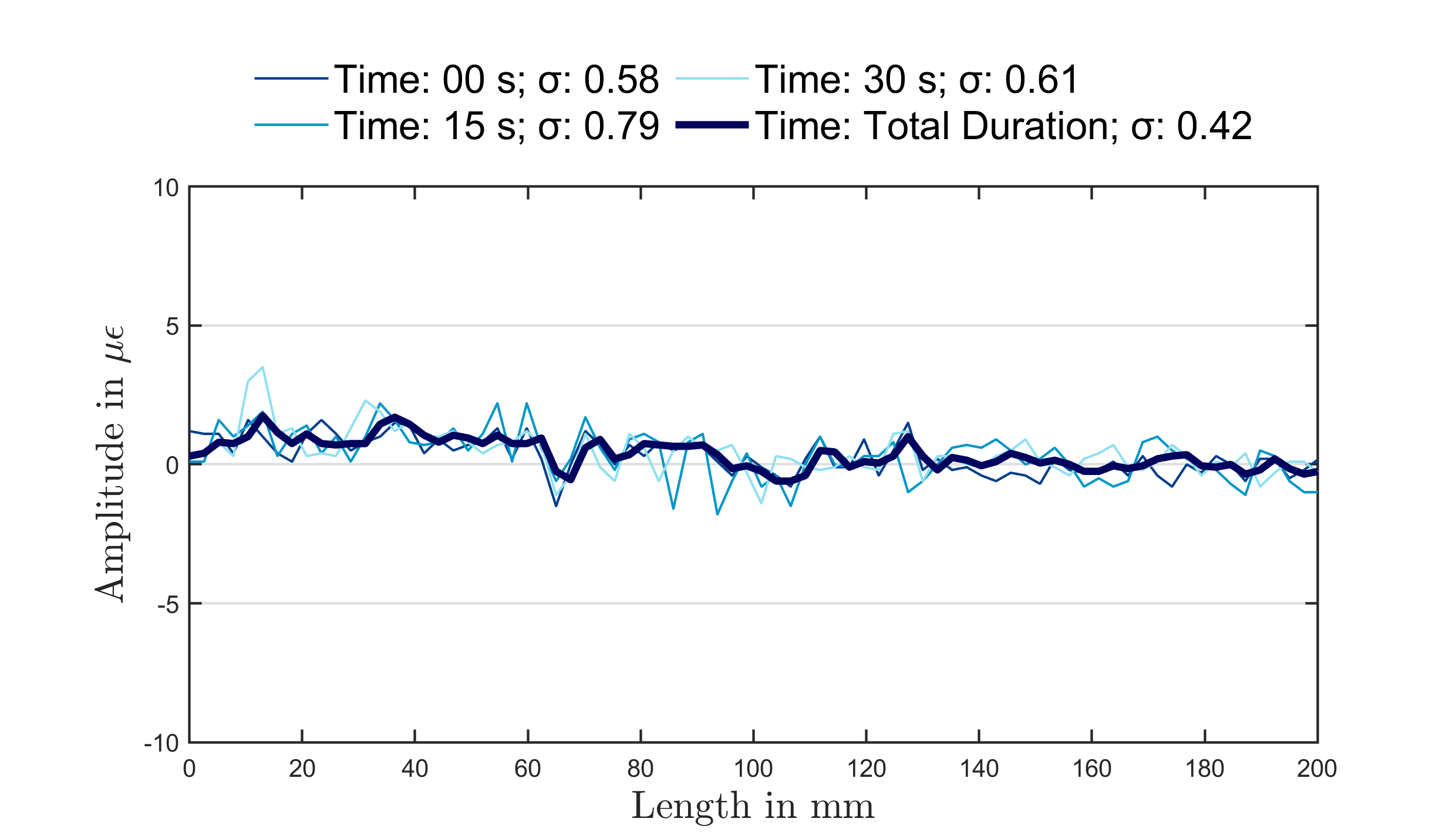}}\hfill%
		\subcaptionbox{\label{fig:ruhemessungmittelwert}}{\includegraphics[width=0.45\textwidth]{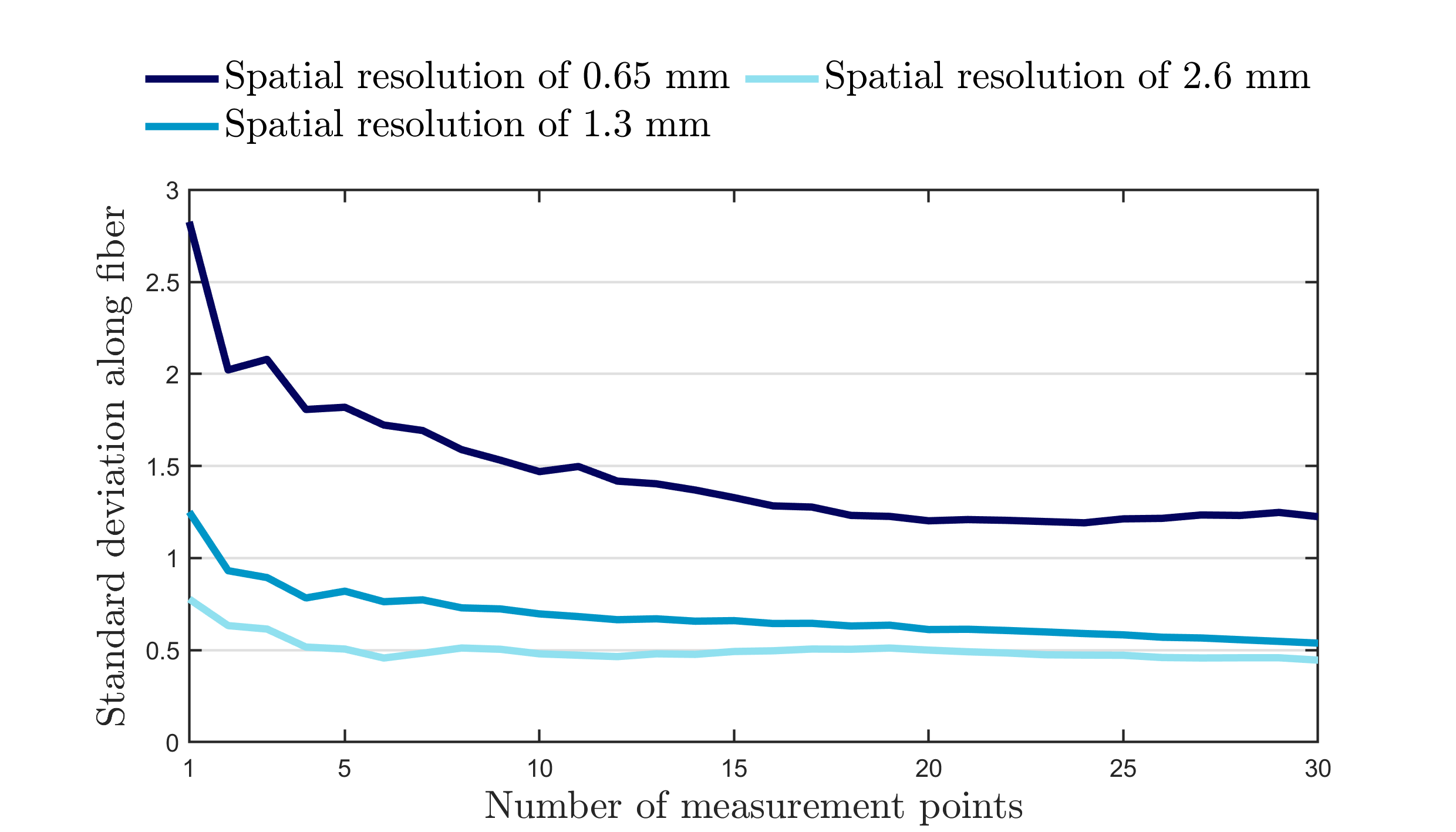}}%
		\hspace*{\fill}%
		\caption{Amplitude of the measurement at different times under homogeneous environmental influences with $\SI{1}{\hertz}$ at different spatial resolutions: $\SI{0.65}{mm}$ (a), $\SI{1.3}{mm}$ (b) and $\SI{2.6}{mm}$ (c). (d) Change in standard deviation under constant test conditions over time at $\SI{1}{\hertz}$ with continuous averaged values at different spatial resolutions is shown.}
		\label{fig:ruhemessung}
	\end{figure}

	The measurement accuracy of fiber optic sensors is influenced by spatial resolution, with higher resolutions often producing noisier signals. As shown in Figure~\ref{fig:ruhemessung065}, measurements at $\SI{0.65}{\mm}$ resolution have standard deviations ranging from $\SI{2.01}{\micro\epsilon}$ to $\SI{2.16}{\micro\epsilon}$, whereas resolutions of $\SI{1.3}{\mm}$ and $\SI{2.6}{\mm}$ yield reduced noise and lower standard deviations (Figures~\ref{fig:ruhemessung13} and~\ref{fig:ruhemessung26}). Furthermore, a correlation was observed between increased spatial resolution and greater measurement variability, with higher spatial accuracy being associated with a more erratic signal. For instance, at $\SI{0.65}{\mm}$ resolution, peak-to-peak fluctuations were recorded at $-\SI{8.2}{\micro\epsilon}$ to $+\SI{6.8}{\micro\epsilon}$, while at $\SI{1.3}{\mm}$ and $\SI{2.6}{\mm}$, the fluctuations decreased to $\SI{5.9}{\micro\epsilon}$ and $\SI{5.3}{\micro\epsilon}$, respectively.
		
	To reduce these fluctuations and improve precision, the data was averaged over 30 consecutive measurements, effectively halving the standard deviation, see Figure~\ref{fig:ruhemessungmittelwert}. The sensor's combined accuracy is specified by the manufacturer at $\SI{\pm 30}{\micro\epsilon}$ (Table~\ref{tab:MeasurementAccuracy}) \cite{lunainnovations_ODiSI_2022}. As the standard deviation is within the expected range for the resolution and based on the test results a spatial resolution of $\SI{2.6}{\mm}$ was selected for subsequent investigations. A lower resolution was deemed acceptable to enhance sensitivity to external influences on the sensor and is therefore considered sufficient for measuring temperature effects in \acrshort{acr:LIB}s. 
	
\begin{figure}[ht]
		\hspace*{\fill}%
		\subcaptionbox{\label{fig:Temperaturort}}{\includegraphics[width=0.45\textwidth]{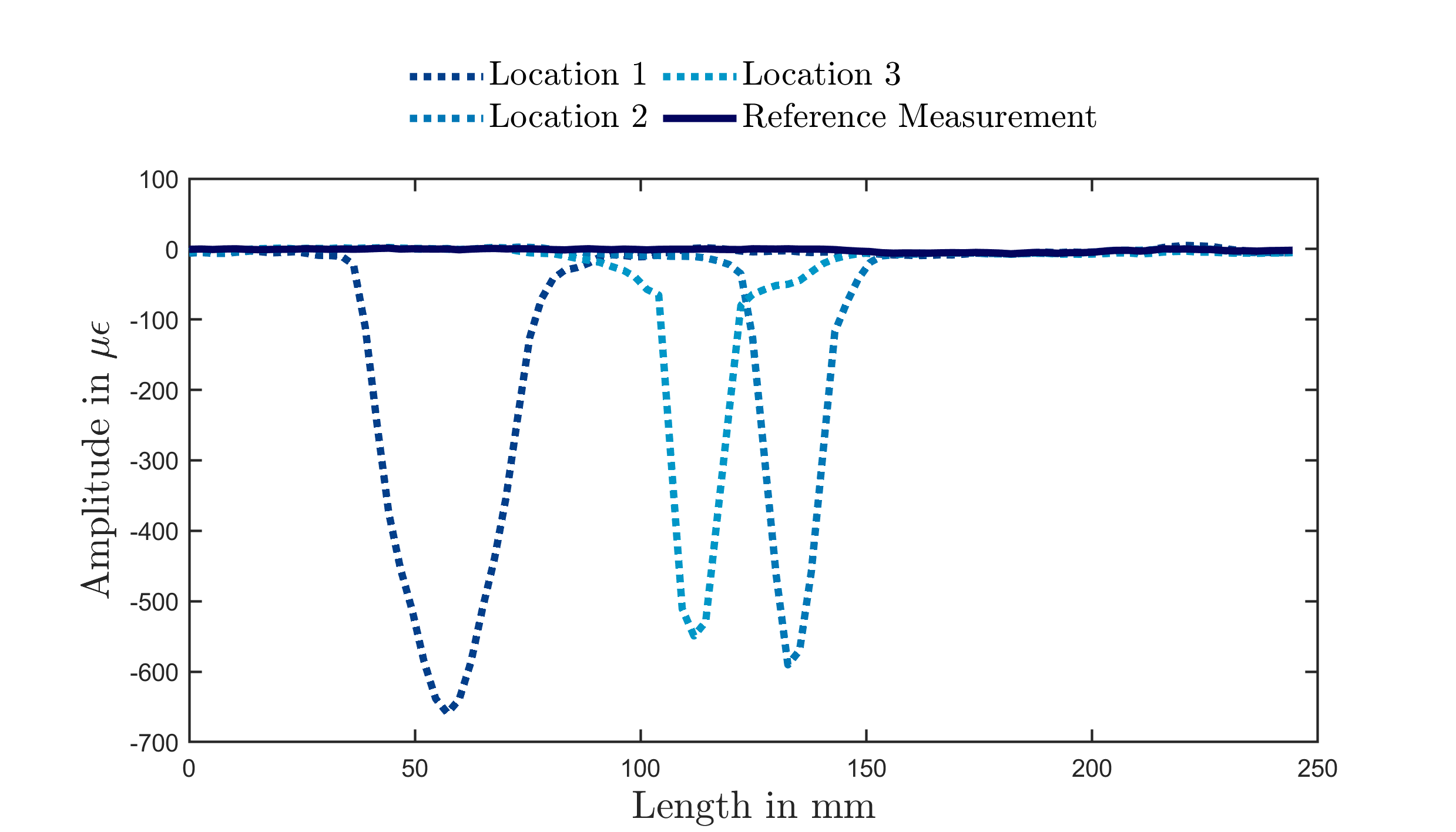}}\hfill%
		\subcaptionbox{\label{fig:Thermalrelaxation}}{\includegraphics[width=0.45\textwidth]{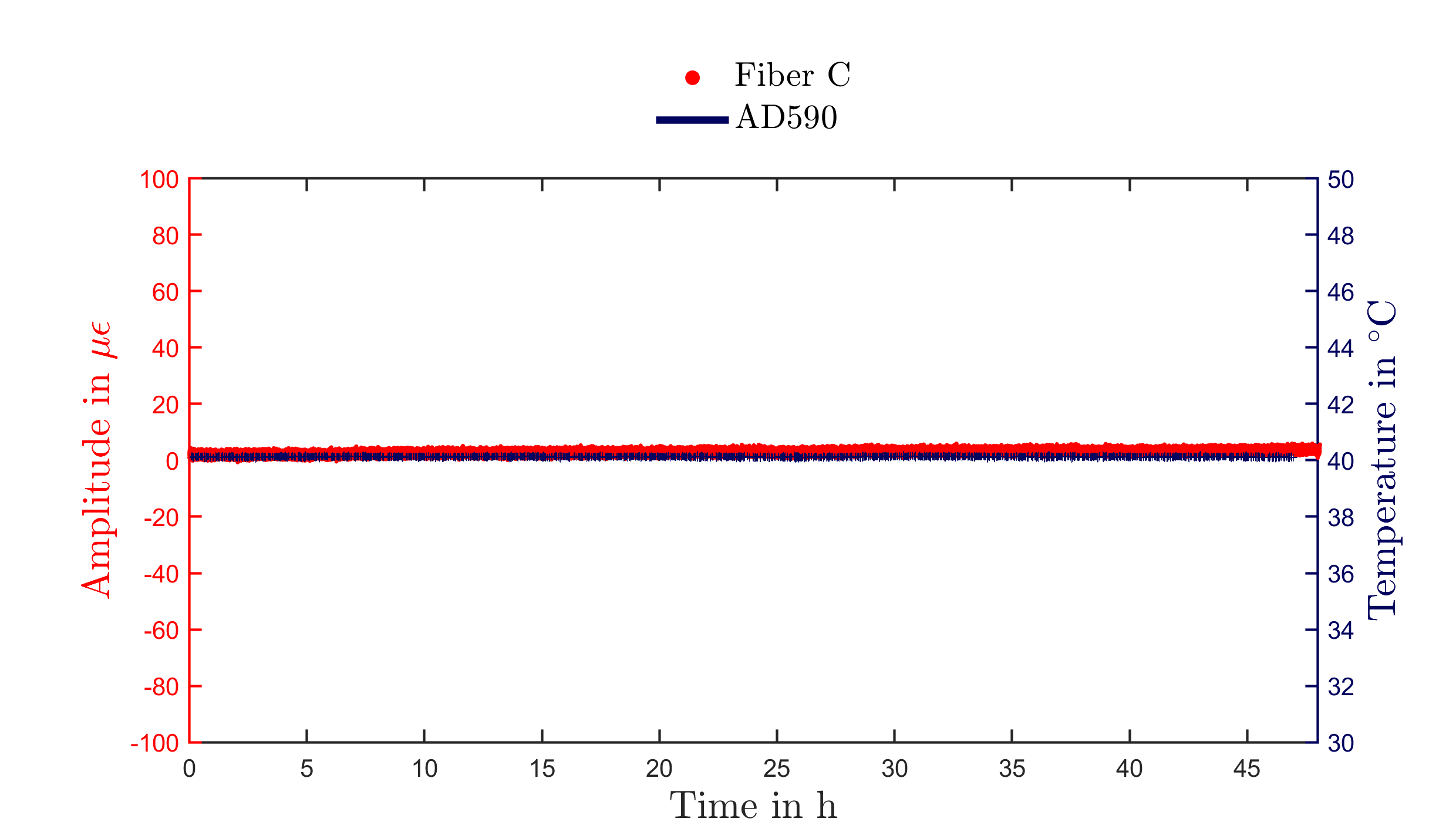}}%
		\hspace*{\fill}%
		\caption{(a) Exemplary measurement of three spatially resolved, time-independent cooling events at $\SI{25}{\degreeCelsius}$ ambient temperature. (b) Relaxation measurement of Fiber C at a constant temperature of $\SI{45}{\degreeCelsius}$ with no pressure applied to the fiber.}
		\label{fig:relax}
	\end{figure}

	As shown in Figure~\ref{fig:Temperaturort}, local cooling events can be clearly identified using a Fiber C, demonstrating the high spatial resolution achievable with optical fibers. Using a \textit{FREEZE 75} cooling spray, the results indicate that a strain of $7-\SI{10}{\micro\epsilon}$ corresponds to a temperature change of approximately $\SI{1}{\degreeCelsius}$, consistent with findings in the literature \cite{kreger_High_2006a,korganbayev_Fiber_2020}. This experiment highlights the sensor's functionality in detecting temperature decreases; however, the principle operates in both directions, meaning the fiber sensor can similarly detect heating events analogously. These strain-temperature relationships might vary depending on the fiber material, coating and other design factors.
	
	The fiber optic sensor’s response accurately matches the amplitude of the cooling event, highlighting its ability to measure temperature variations with high spatial precision. The subsequent experiments, based on the previously defined parameters, explore different fiber types under various configurations, with temperature and positioning being examples of the factors varied to assess their performance under different conditions. In the following, we focus on a specific section of the fiber and examine a series of tests to quantify the temperature dependency of the individual fiber types presented.
	\newline
	
	\textbf{Thermal Relaxation}
	\newline
	
	During the 48-hour measurement period described in the earlier experimental setup, the thermal relaxation behavior of the fiber sensor was investigated. The sensor exhibited a maximum deviation of $5 \, \mu\epsilon$, as shown in Figure~\ref{fig:Thermalrelaxation}. A comparison of the sensor's output at the start and end of the measurement revealed a drift of approximately $1 \, \mu\epsilon$. However, this minor sensitivity cannot conclusively be attributed to thermal drift. The observed change remains within the specified accuracy limits of both the measuring device and the sensor. Since the observed deviations remained within the spatial and temporal accuracy limits of the reference sensors and the homogeneity of the test chamber, no evidence of thermal relaxation was observed. This consistency persisted throughout the entire duration of all experiments, underscoring the stability of the used fiber sensors for extended periods.
	\newline
	
	\clearpage
	
	\textbf{Temperature Dependence}
	\newline

	\begin{figure}[ht]
		\subcaptionbox{\label{fig:3Dplot}}{\includegraphics[width=0.45\textwidth]{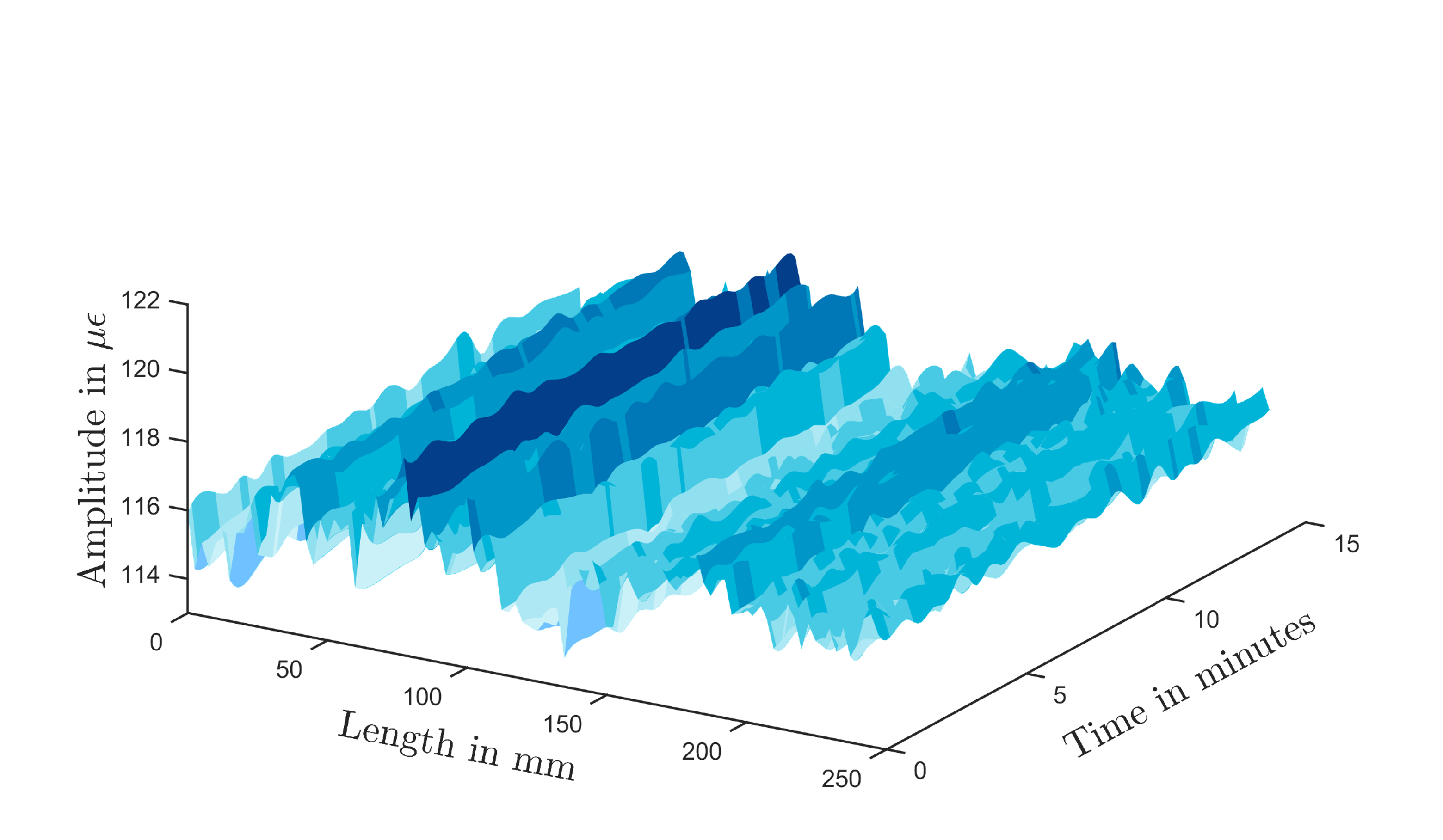}}\hfill%
		\subcaptionbox{\label{fig:TemperaturVergleichFaserLänge}}{\includegraphics[width=0.45\textwidth]{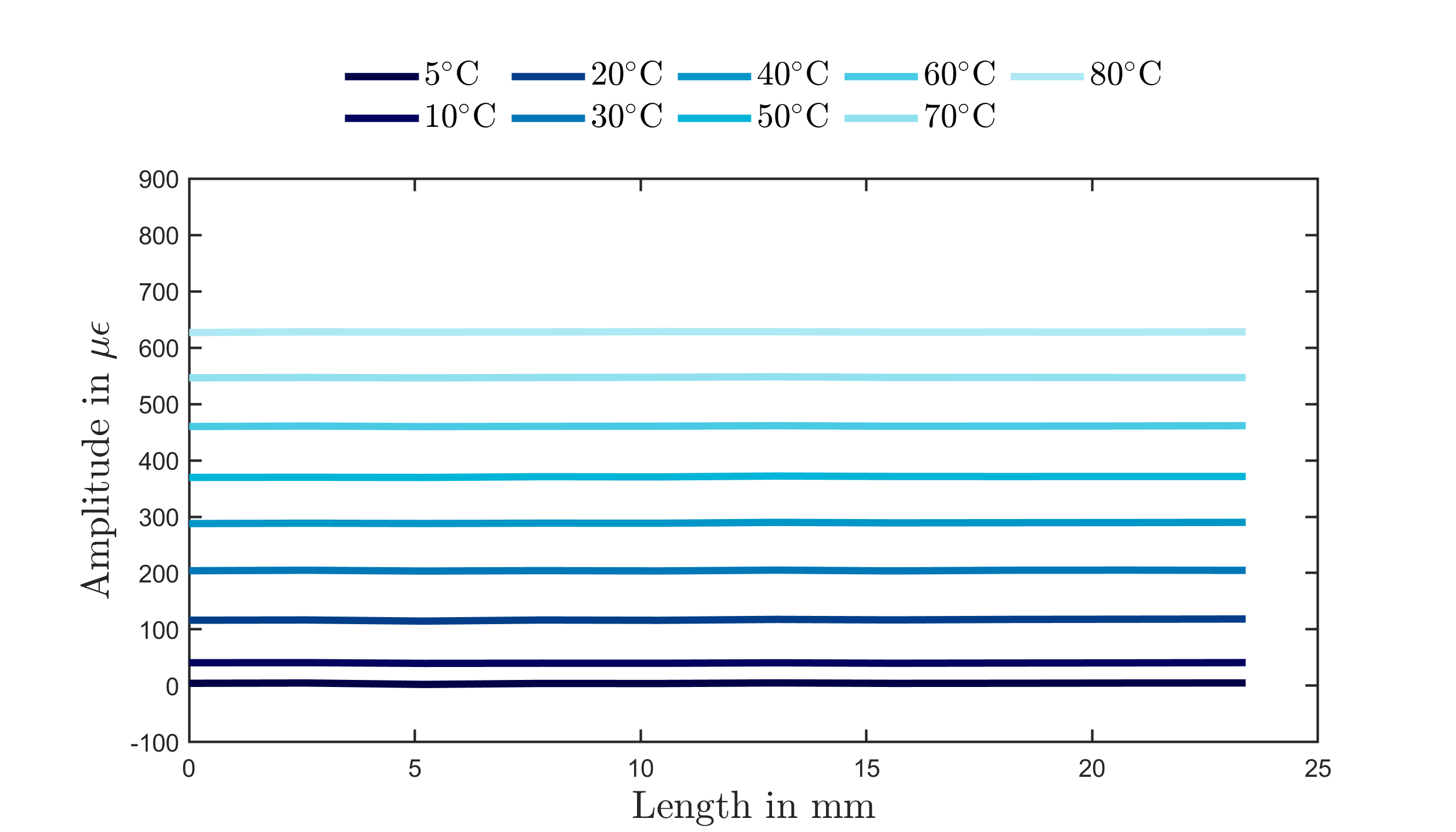}}%
		\hspace*{\fill}%
		
		\smallskip
		
		\hspace*{\fill}%
		
		\subcaptionbox{\label{fig:TemperaturVergleich}}{\includegraphics[width=0.45\textwidth]{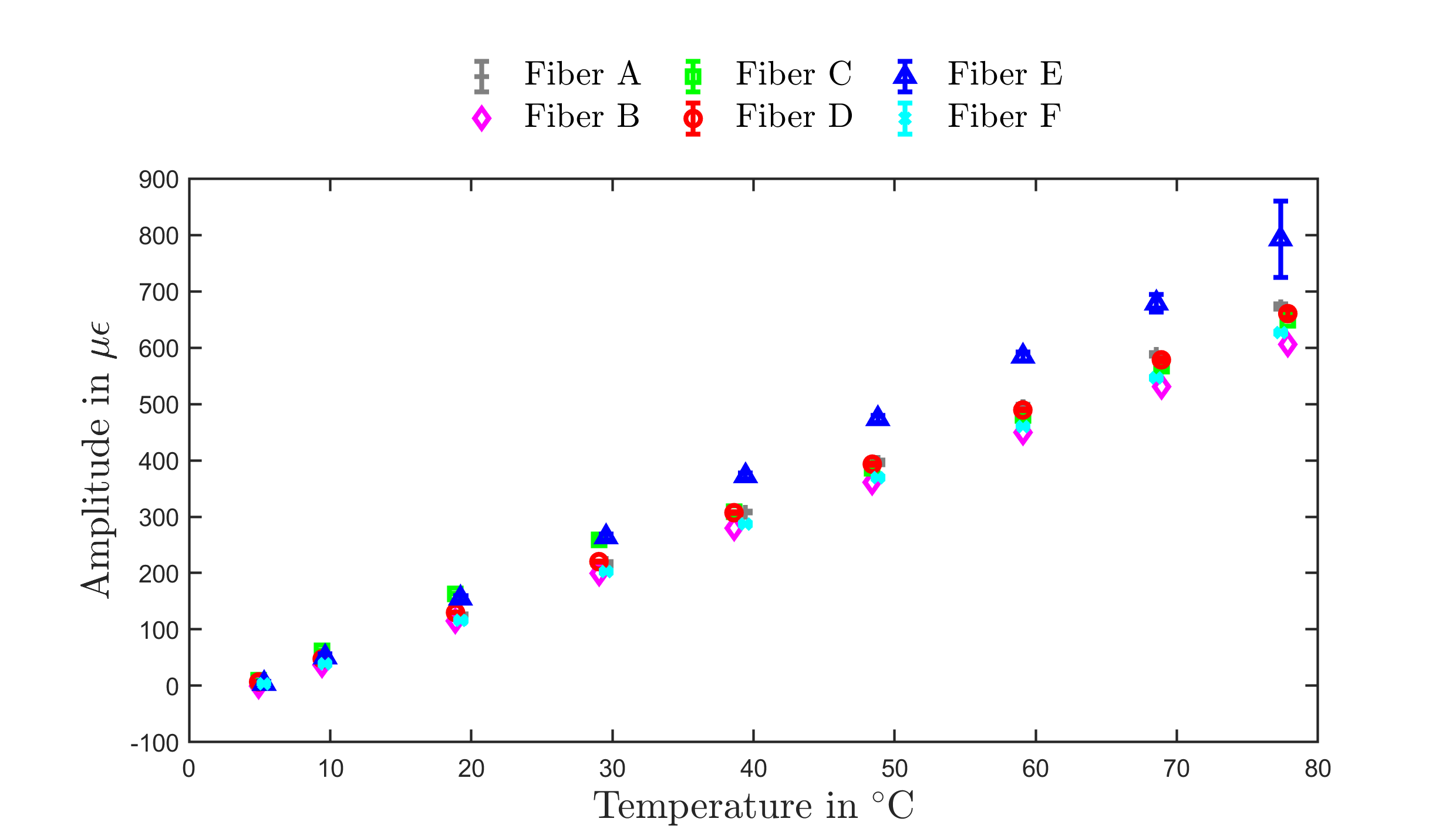}}\hfill%
		\subcaptionbox{\label{fig:TemperaturFit}}{\includegraphics[width=0.45\textwidth]{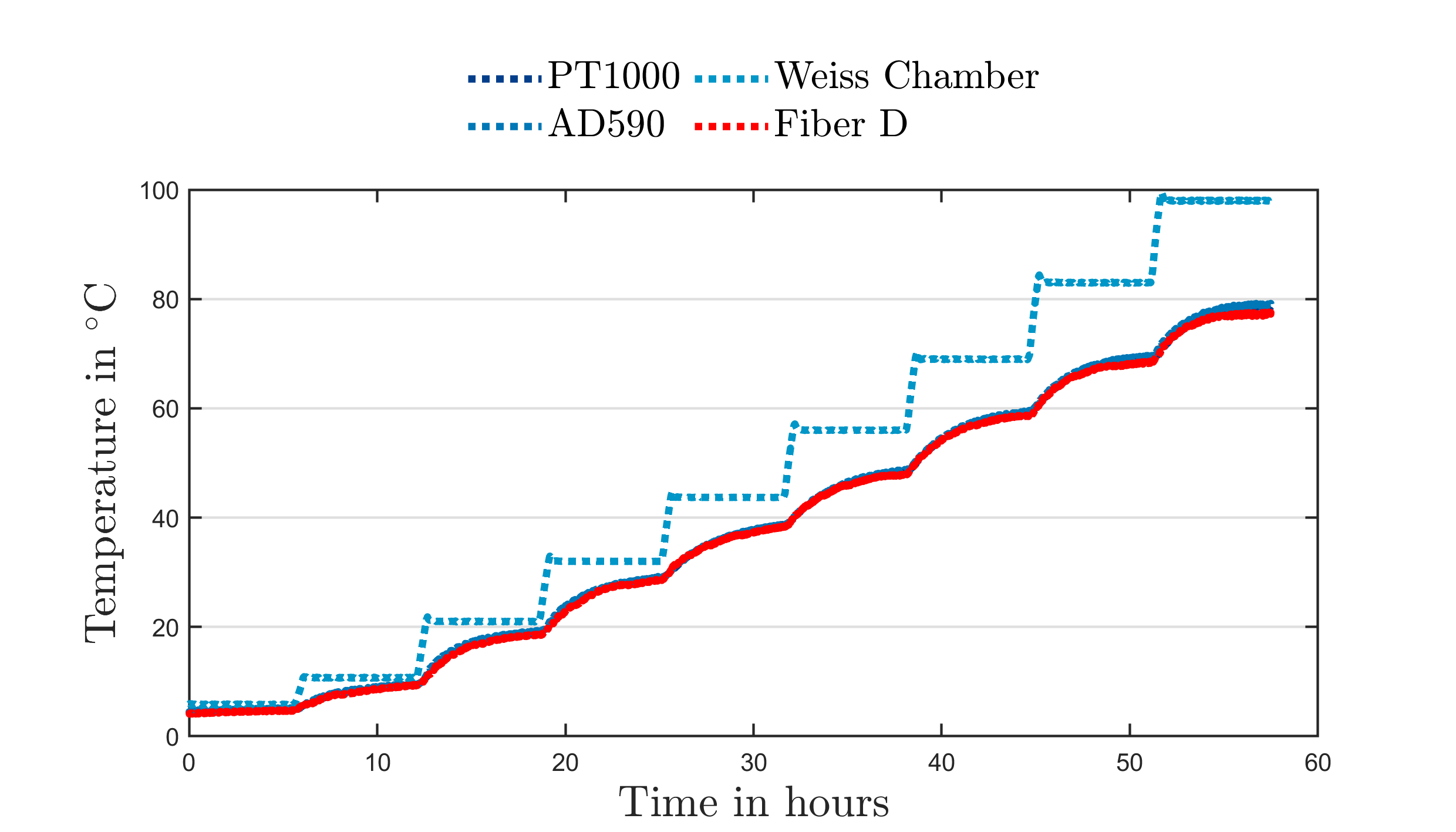}}%
		\hspace*{\fill}%
		
		\caption{(a) Example of a 3D spatial resolution of measurement signal over measurement duration and length of the sensor (Fiber F) at $\SI{20}{\degreeCelsius}$. (b) Example of a time-averaged measurement of a sensor section (Fiber F) across different temperatures. (c) Temperature dependency of various fiber types. (d) Stepwise temperature increase in the chamber with reference sensor and a fitted measurement of Fiber D with the calculated coefficient from (c).}
		\label{fig:}
	\end{figure}

	In Figure~\ref{fig:3Dplot}, an example of the three-dimensional resolution of a measurement over a 15-minute duration at $\SI{20}{\degreeCelsius}$ is shown. The thermal stress profile along the $\SI{250}{\mm}$ length of the fiber displays a similar pattern, with a maximum delta between the minimum and maximum of about $\SI{5}{\micro\epsilon}$. In Figure~\ref{fig:TemperaturVergleichFaserLänge}, the same trend is depicted across other temperature levels, where a time-averaged mean was taken over the 15-minute measurement period. These results along the fiber are consistent, with no noticeable outliers or significant variation. This observation confirms the uniformity of the spatial resolution along the fiber, as initially assumed, considering the defined constraints in the previous section.
		
	Figure~\ref{fig:TemperaturVergleich}, illustrates  the temperature dependencies of the individual fiber types. In this case, the measured data, as shown in Figure~\ref{fig:TemperaturVergleichFaserLänge}, was additionally averaged over the length and represented according to their variance. For Fiber E, an increase in the variance of the measurements with rising temperature was observed, which was noticeable compared to the other fiber types. The deviation here was in the range of $\pm \SI{50}{\micro\epsilon}$, while for the others, it was at most $\pm \SI{7}{\micro\epsilon}$. Presumably, the fiber was no longer fully submerged in the water upon reaching $\SI{80}{\degreeCelsius}$, leading to a higher measuring deviation compared to the other measurements. This, however, could not be definitively confirmed. The measurement enables the determination of the coefficients $\mi{K}{T}$, as presented in Table~\ref{tab:Koeffizient}. As shown, the fibers demonstrate a quasi-linear behavior across the tested temperature range. These coefficients can be used to derive the temperature, as illustrated in Figure~\ref{fig:TemperaturFit}. With a maximum deviation of $\pm \SI{2}{\degreeCelsius}$, this falls within the range of possible accuracy for both the reference measurements and the fiber data. Despite differences in core doping and coating, the temperature-induced length change behavior of the fibers is similar to literature values in the range reported earlier.
	
	\begin{table}[ht]
	\caption{The temperature coefficients $\mi{K}{T}$ of the fibers used in this paper were determined as follows:}
	\footnotesize
	\centering
	\begin{tabular}{ll}
		\toprule
		ID & Coefficient $\mi{K}{T}$ in $\SI{}{\frac{\mu\epsilon}{\degreeCelsius}}$ \\ 
		\midrule
		Fiber A & 9.2779 \\
		Fiber B & 8.3102 \\ 
		Fiber C & 8.4613 \\ 
		Fiber D & 8.9459 \\
		Fiber E & 10.8307 \\
		Fiber F & 8.6413 \\
		\bottomrule
	\end{tabular}
	\label{tab:Koeffizient}
	\end{table}
	
	In Figure~\ref{fig:TemperaturFit}, the data from the two reference sensors, the climate chamber temperature and the fitted measurement for Fiber D, derived using the calculated coefficient, are presented. The chamber's temperature shows a brief transient process before reaching the set value. The water basin's temperature, as it attempts to follow the chamber's temperature, is temporally delayed due to its thermal mass and associated inertia until a homogeneous state is achieved. As evident, the ambient temperature of the chamber differs with rising temperature from the measured one in the water basin. This difference can partly be attributed to the evaporation of water and its associated cooling effect. 
	Furthermore, it appears that the water basin reaches a steady state at each temperature plateau, which deviates from the chamber's temperature. This deviation is likely due to the water's thermal inertia and insufficient heat transfer. However, given the use of reference sensors for temperature measurement, this deviation is not considered critical.
	\newline
	
	\textbf{Temperature Dependence in Bending Radii}
	\newline

	\begin{figure}[ht]
		\hspace*{\fill}%
		\subcaptionbox{\label{fig:TemperaturRadiushFaserLänge}}{\includegraphics[width=0.45\textwidth]{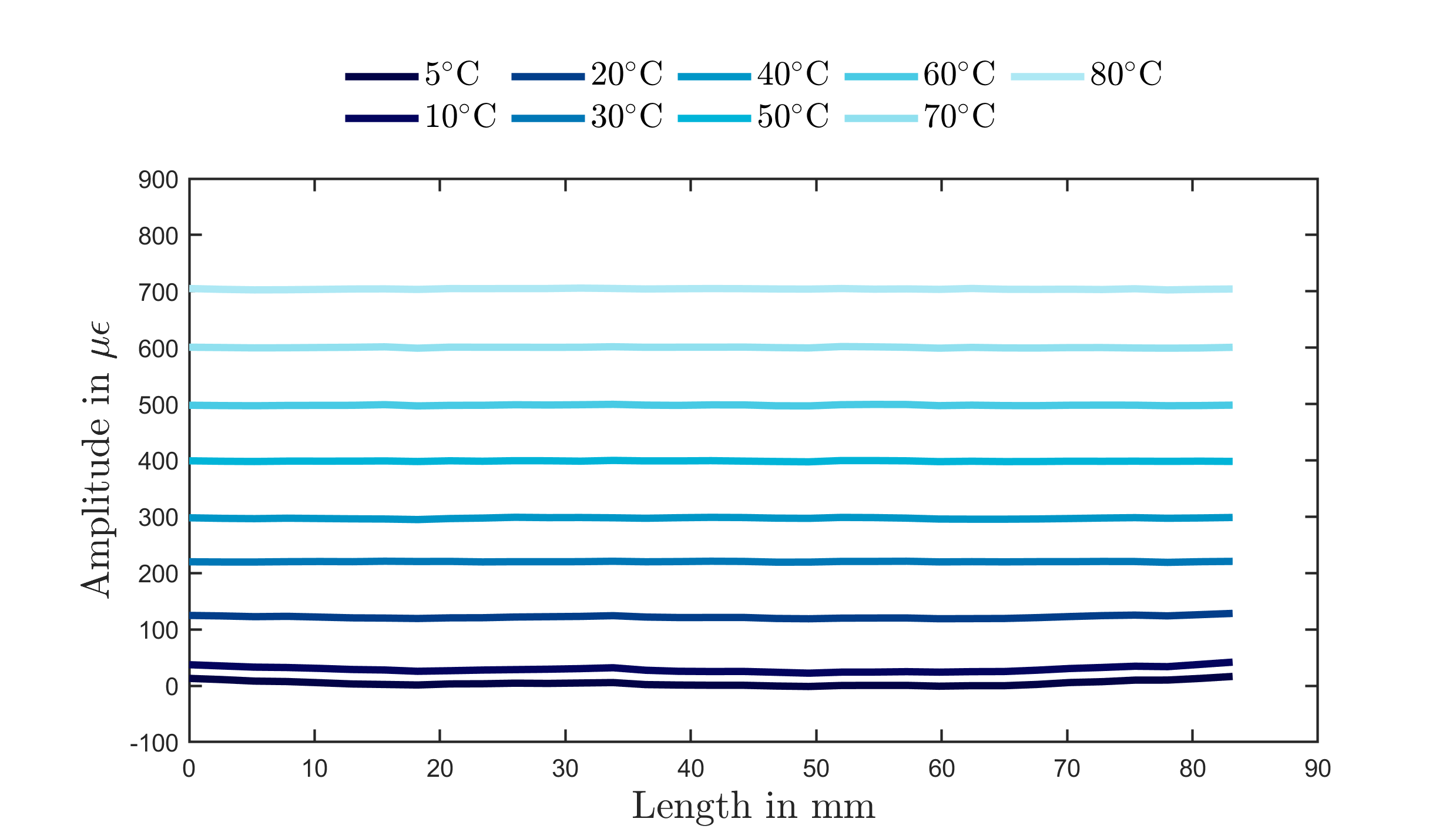}}\hfill%
		\subcaptionbox{\label{fig:TemperaturRadius}}{\includegraphics[width=0.45\textwidth]{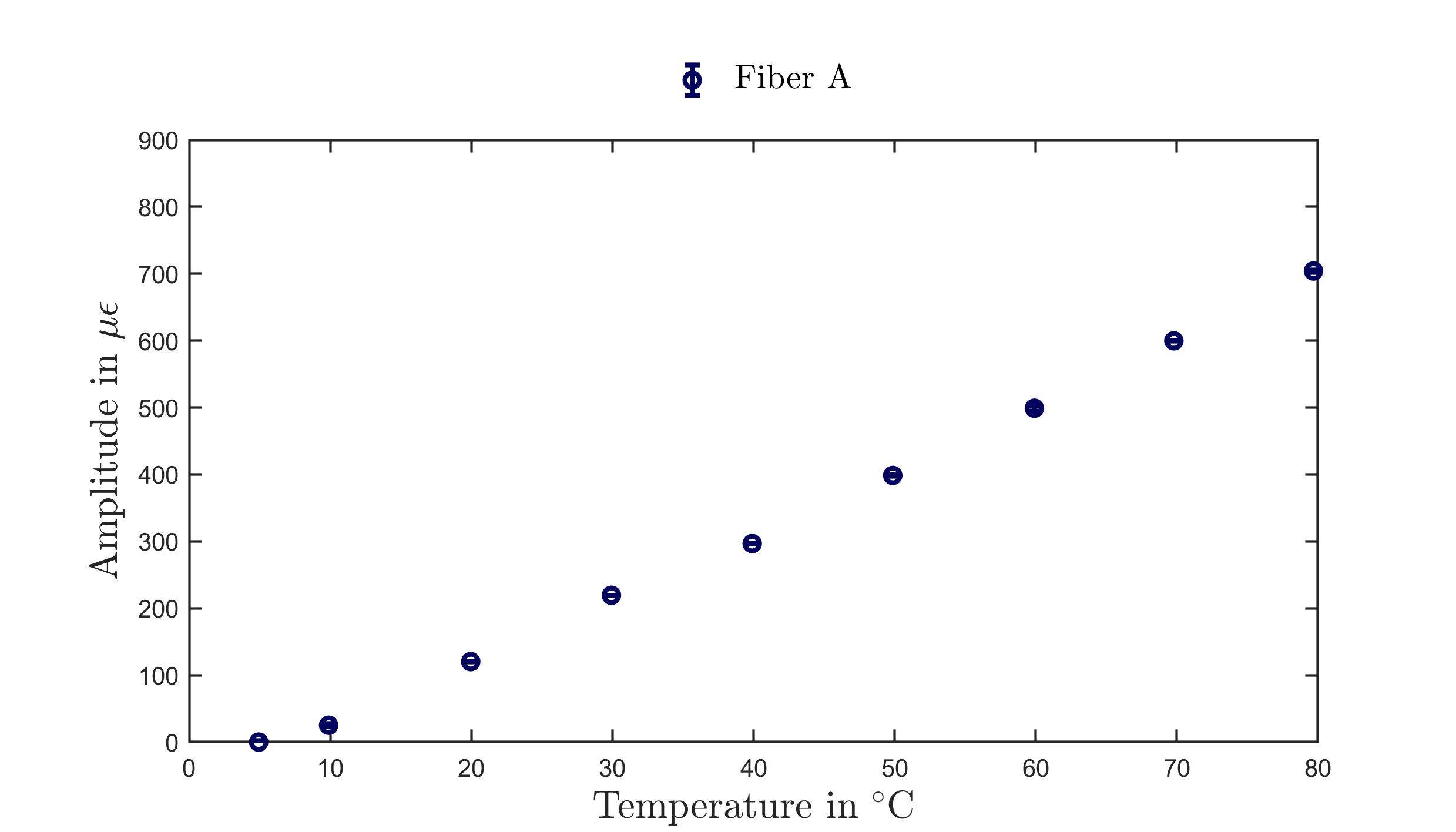}}%
		\hspace*{\fill}%
		\caption{(a) Time-averaged measurements of sensor segments in bends with varying radii of Fiber A across a temperature range from $5$ to $\SI{80}{\degreeCelsius}$. (b) Temperature dependence along the length of Fiber A within the tightest bend (radius of $\SI{15}{\mm}$).}
	\end{figure}
	
	Examining the effect of varying radii on the fiber’s response across different temperature levels, it can be observed that the measurements remain consistent over the entire length of the fiber, particularly within the smallest radius, as shown in Figure~\ref{fig:TemperaturRadiushFaserLänge}. Both the beginning and end of the fiber, which include short straight sections, are also part of the measurement, indicating that the radii do not significantly affect the results. To determine a valid coefficient for the fiber used in this setup, the spatial average of the measured values at the different radii across the temperature levels was calculated. Figure~\ref{fig:TemperaturRadius} illustrates the relationship between temperature and amplitude, with error bars, while also factoring in the variations in radii. As clearly shown, the variance is close to zero.
	
	No significant difference in the measurement values between different radii in the range of $15$ to $\SI{60}{\mm}$ was detected, as the maximum deviation observed was only $\SI{2.16}{\mu\epsilon}$. Additionally, it was demonstrated that the sensor's position, whether along a radius or on a straight path, had negligible impact on temperature sensitivity. The difference between the radius and straight path measurements was only a maximum of $\SI{ 0.273}{\mu\epsilon}$ at a $\SI{20}{\mm}$ radius and $\SI{20}{\degreeCelsius}$. A coefficient of $\mi{K}{T} = \SI{9.017}{\frac{\mu\epsilon}{\degreeCelsius}}$ was determined for this case using linear regression. This value is close to the reference measurement performed and described in the previous section.

	\section{Impact of Specific Steps during Battery Production on the Functionality of the Sensor}
	\label{kap:Samples}
	
	Inserting a fiber requires certain steps that could influence the measurement signal, such as precise alignment, ensuring minimal stress on the fiber and avoiding any potential interference from the surrounding materials or manufacturing processes. In the following, we focus on the effects of sealing seams of a pouchbag and immobilisation of the fiber on the measured signal. 	
	
	As previously mentioned, several attempts to integrate fiber optic sensors into cells have been described in the literature, but most of these were limited to insertion into round cells. The integration of the fiber into the cell is a non-trivial process, requiring careful handling to ensure proper placement. While the procedure allows for flexibility in positioning, it involves certain challenges in terms of precision and alignment. One approach is to insert the fiber after the production of the cell, which can be achieved by simply opening the pouch bag and carefully placing the fiber inside. However, this approach restricts the insertion of the sensor to the outer ends of the cell stack, limiting the ability to monitor the temperature across the entire stack effectively.
		
	\begin{figure}[h!]
		\centering
		\subcaptionbox{\label{fig:PouchSample1}}{\includegraphics[width=0.9\textwidth]{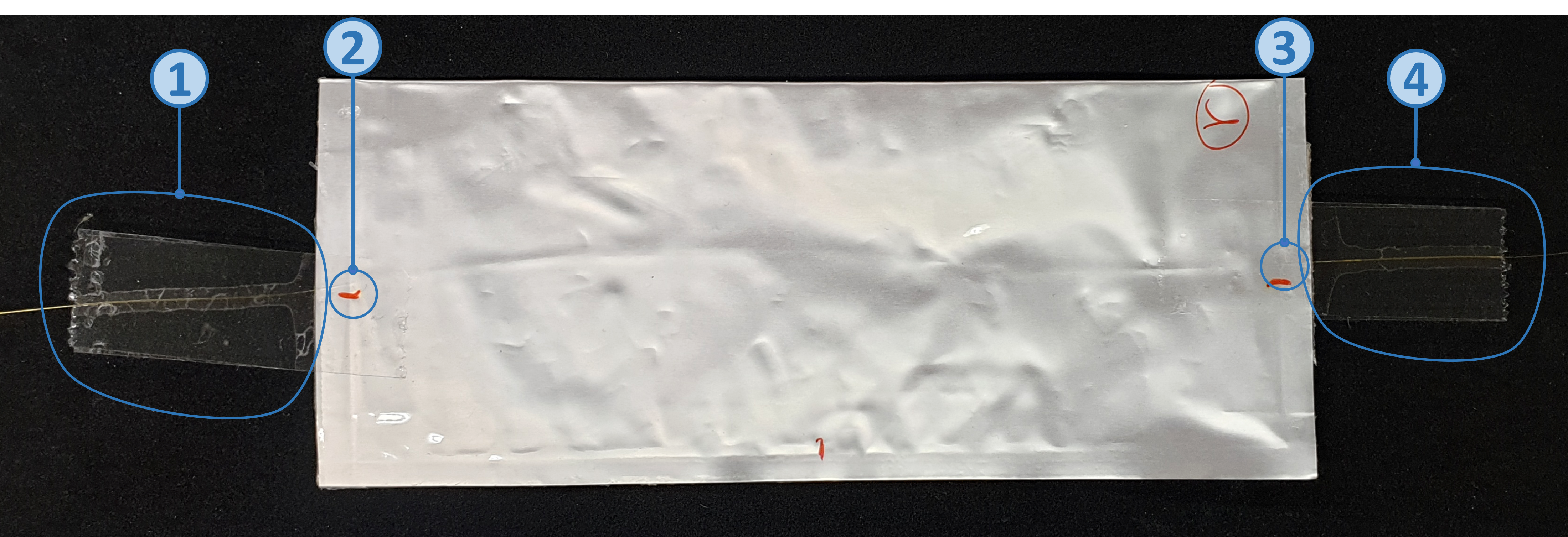}}
		\smallskip
		\subcaptionbox{\label{fig:PouchSample1_T}}{\includegraphics[width=0.9\textwidth]{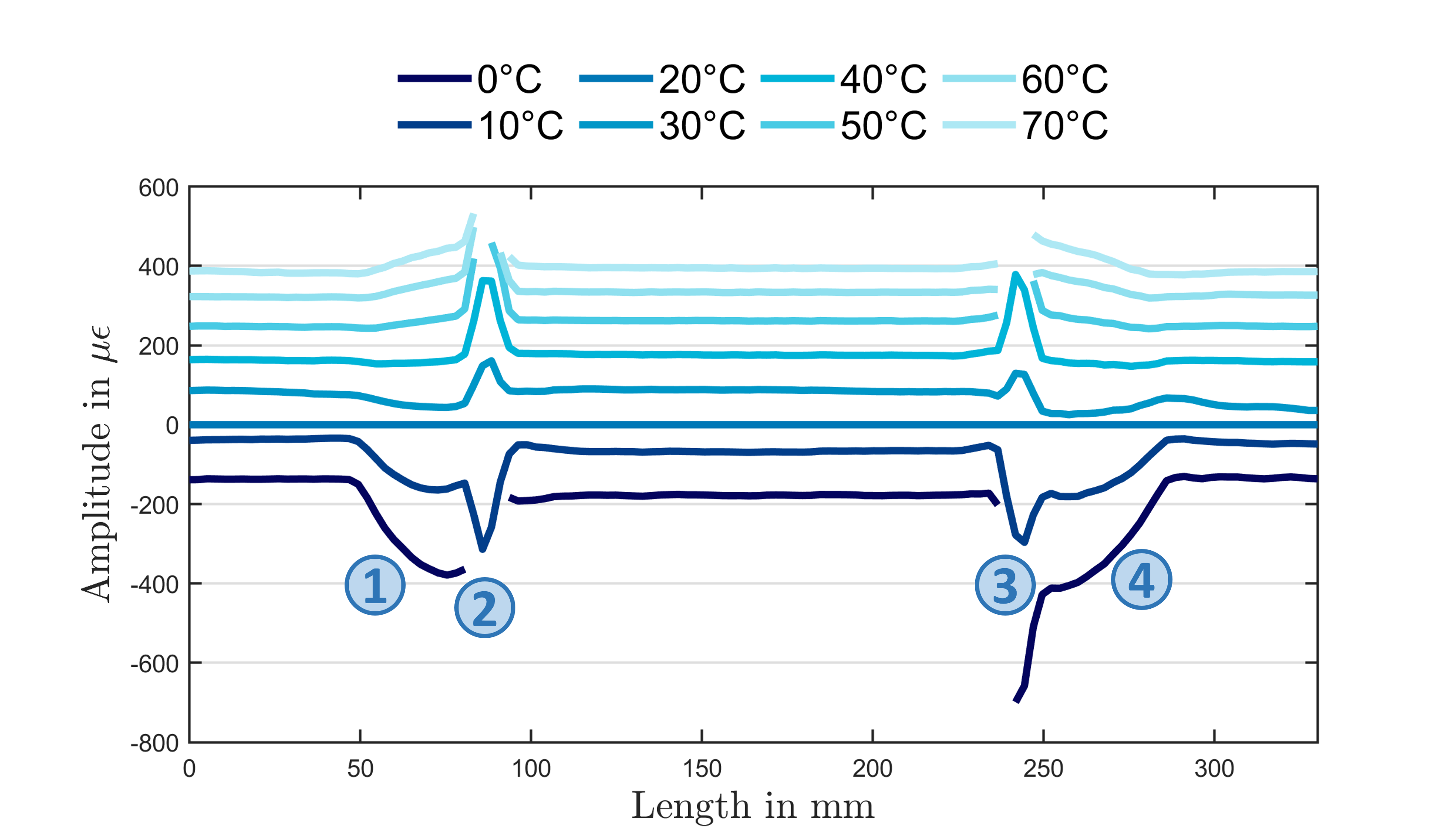}}
		\caption{(a) Sealed pouch bag with integrated Fiber B sensor with adhesive tape attachments. (b) Corresponding measurement results of the thermal changes of the sample from $0$ to $\SI{70}{\degreeCelsius}$ (zero-point calibration at $\SI{20}{\degreeCelsius}$).}
		\label{fig:Pouch1}
	\end{figure}	
	
	For more accurate measurement of internal cell temperature, it may be more beneficial to integrate the fiber directly into the cell stack during the production process. To increase the measurement area, it is also advantageous to place the fiber not only in a straight line, but also in bends, thus covering a larger area within the battery cell. When integrating fiber optic sensors into pouch cells, it is crucial to guarantee complete sealing of the cell for both functionality and safety of the cells. A poor sealing of the cell can very quickly lead to unwanted leakage and damage.  
	
	\begin{figure}[h!]
		\centering
		\subcaptionbox{\label{fig:PouchSample2}}{\includegraphics[width=0.9\textwidth]{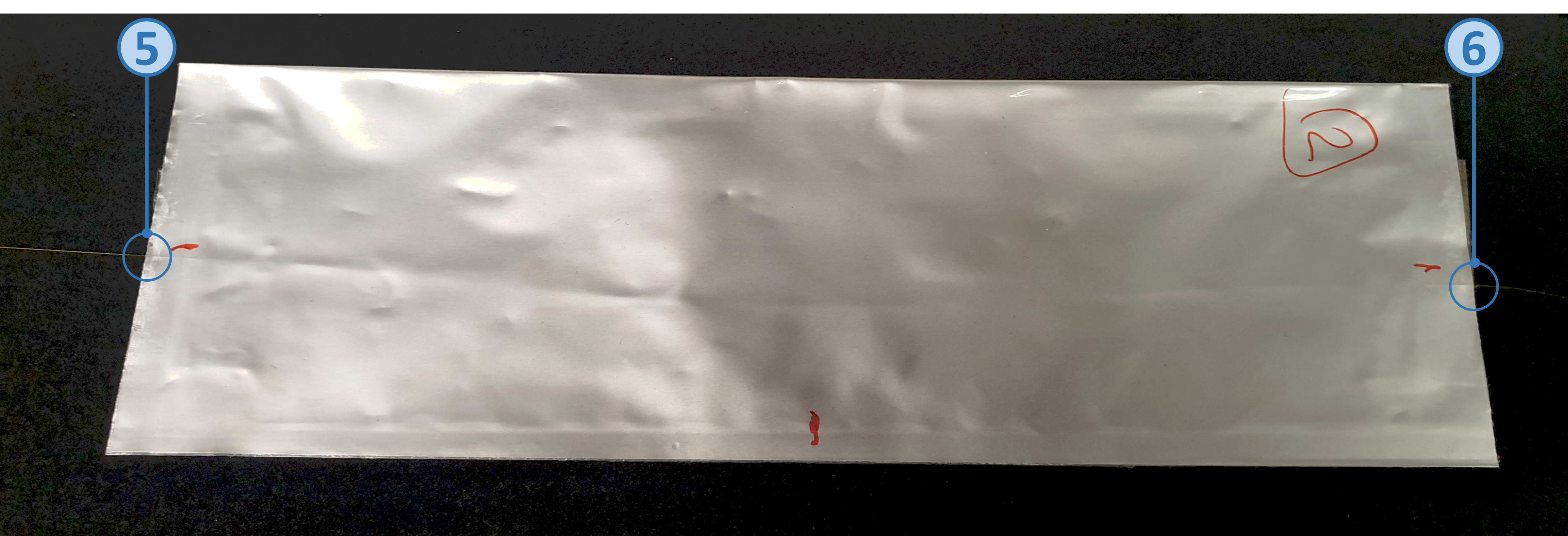}}
		\smallskip
		\subcaptionbox{\label{fig:PouchSample2_T}}{\includegraphics[width=0.9\textwidth]{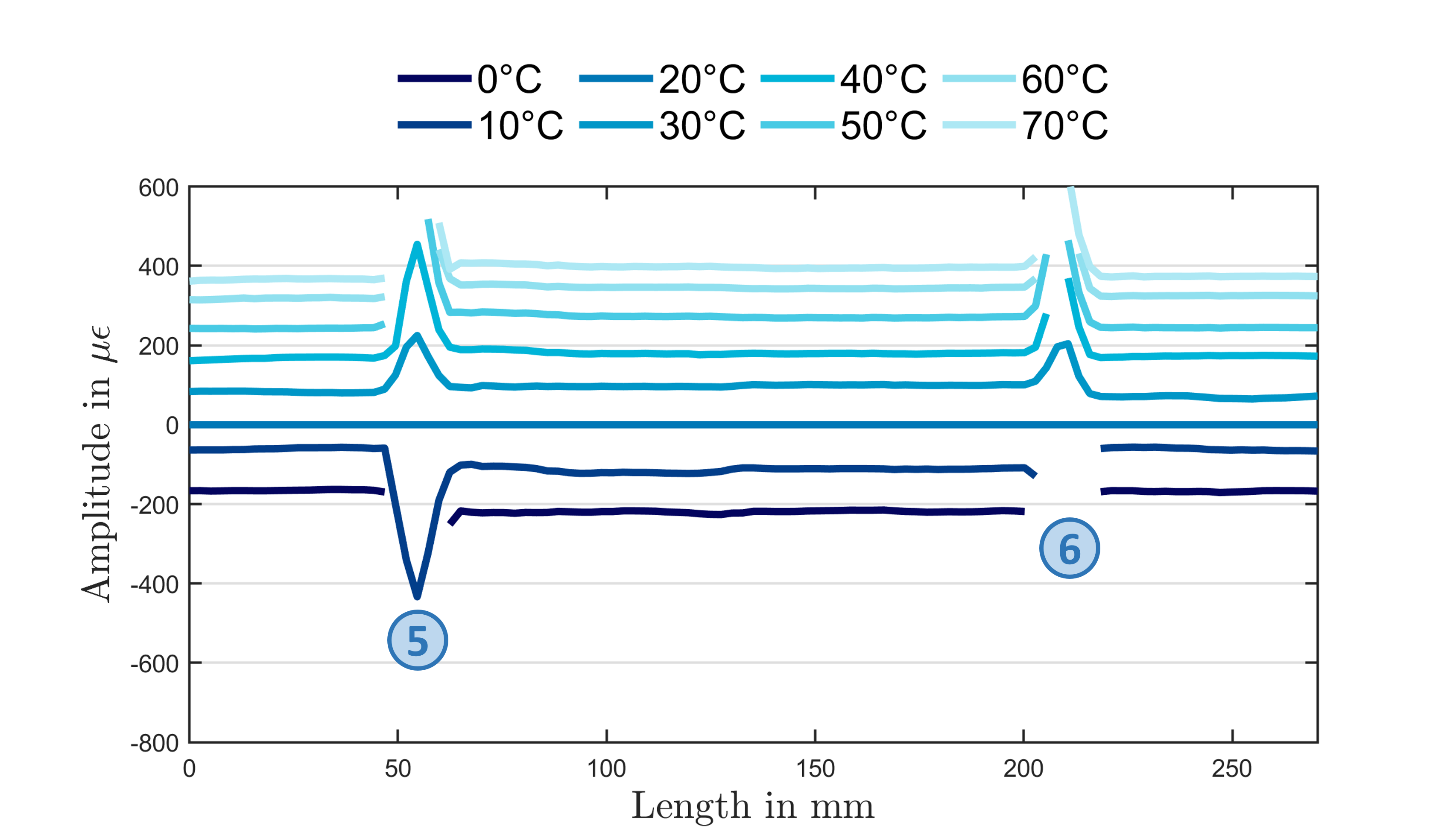}}
		\caption{(a) Sealed pouch bag with integrated Fiber B sensor without adhesive tape attachments. (b) Corresponding measurement results of the thermal changes of the sample from $0$ to $\SI{70}{\degreeCelsius}$ (zero-point calibration at $\SI{20}{\degreeCelsius}$).} 
		\label{fig:Pouch2}
	\end{figure}	
	
	On various samples a fiber was inserted into a pouch bag, sealed and then processed into a sensor. Figure~\ref{fig:PouchSample1} and Figure~\ref{fig:PouchSample2} show two pouch bag samples, as used for lithium-ion pouch cells, containing an inserted fiber sensor of type Fiber B with the external dimension of $\SI{165}{\mm}$ x $\SI{70}{\mm}$ (LxW). 
	
	To better protect the fiber optic sensor, an adhesive strip was applied to the outer surface of the pouch bag in the setup shown in Figure~\ref{fig:PouchSample1}. This was done to prevent the sensor from potential damage due to movement or mechanical stress, especially near the seal seam. However, to assess the effect of this external protection, a second experiment was conducted without the adhesive strip, as shown in Figure~\ref{fig:PouchSample2}, to observe how much the adhesive actually influences the sensor's performance. In this experiment, samples were subjected to identical conditions, with temperatures ranging from $\SI{0}{\degreeCelsius}$ up to $\SI{70}{\degreeCelsius}$ in a climate chamber, as previously described.
	
	The measurement results in Figure~\ref{fig:PouchSample1_T} show the entire fiber routing inside the pouch bag, as well as some part of the adjacent sections of the outer fiber visible in the Figures~\ref{fig:PouchSample1}. The maxima and minima along the fiber measurements indicate the areas of the sealing (\circled{2} and \circled{3}) and additional adhesive (\circled{1} and \circled{4}) strip of the sensor, respectively. The fiber sensors were tared at $\SI{20}{\degreeCelsius}$. 

	As can be seen, the measured signal can no longer be displayed in areas of the sealing $\circled{2}$ and $\circled{3}$ as the temperature increases or decreases. This is clearly visible in Figure~\ref{fig:PouchSample1_T}, as well as in Figure~\ref{fig:PouchSample2_T}, particularly in the areas marked by $\circled{5}$ and $\circled{6}$. This effect is due to the internal algorithm of the interrogator and the exceeding of the measuring range, due to high, locally superimposed pressure. This is also clearly visible in the measurement gaps. After the experiment, the seal seam was intentionally opened, but subsequent optical inspection revealed no damage to the coatings or fibers. 
	As expected, the graphical evaluations of the measurements demonstrate that separate measurements conducted on the prepared samples yield nearly identical results. These demonstrate uniform behavior along the non-stressed part of the fiber, with a maximum deviation of $\pm\SI{6}{\micro\epsilon}$, which is in the range of the already reported average standard deviation for temperature measurements. Despite the presence of local artifacts at the seal seams, similar temperature dependencies with an average of $\mi{K}{T} = \SI{8,17}{\frac{\micro\epsilon}{\degreeCelsius}}$ can be observed across the entire measured temperature range. This leads to the conclusion that despite the obvious influence of the sealing and tape on the measurement signal, a valid temperature measurement along the fiber is possible.
	
	\begin{figure}[ht]
	\centering
		\subcaptionbox{\label{fig:AnodeSample1}}{\includegraphics[width=0.9\textwidth]{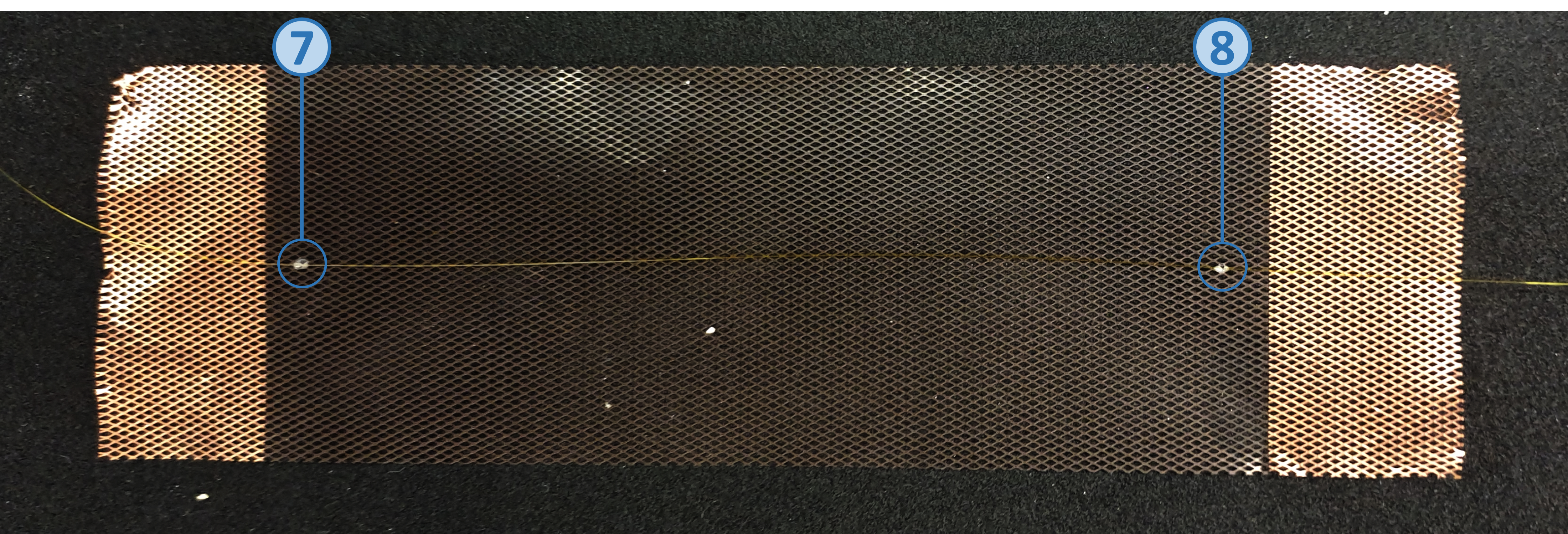}}
		\smallskip
		\subcaptionbox{\label{fig:AnodeSample1_T}}{\includegraphics[width=0.9\textwidth]{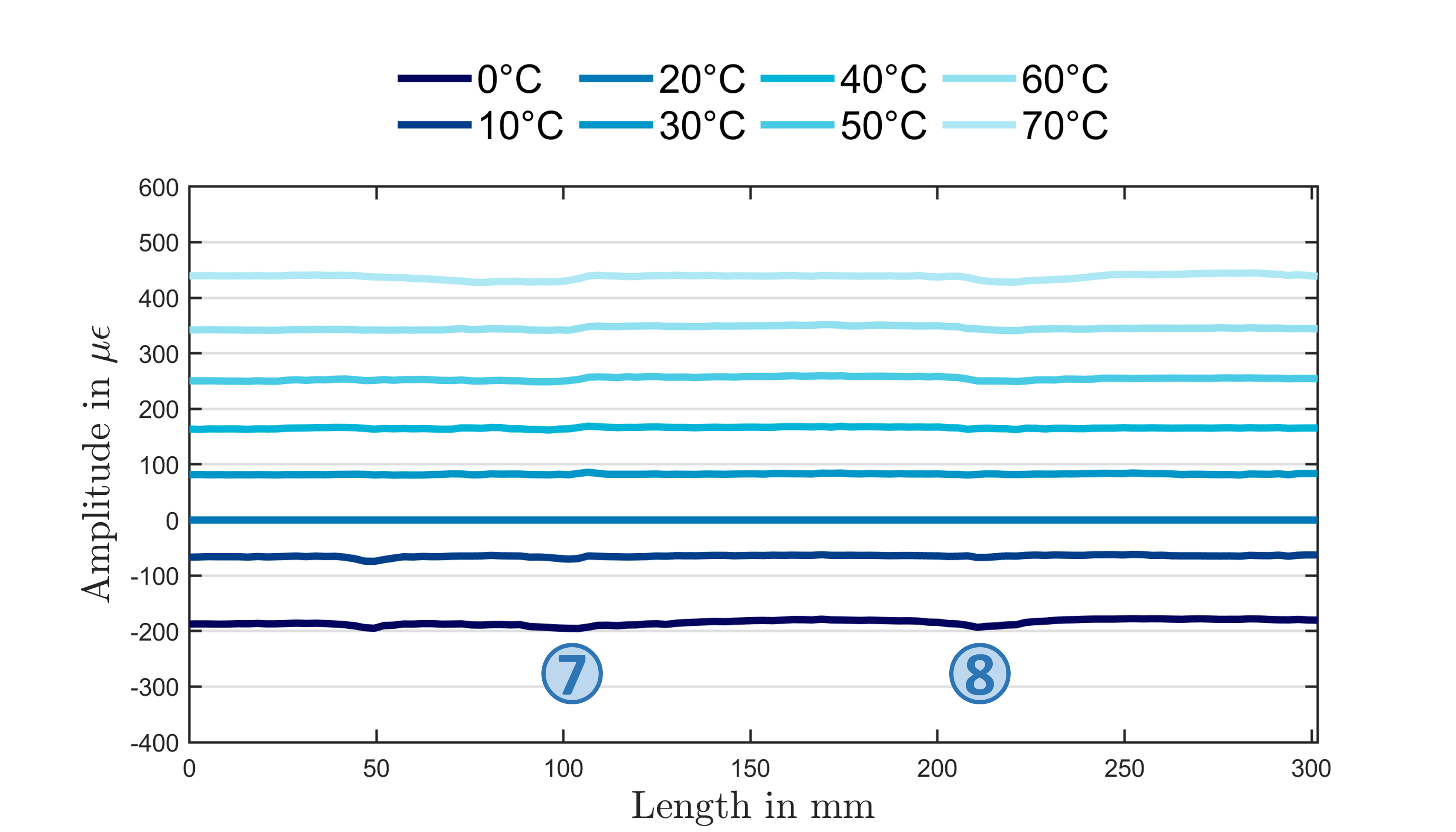}}
		\caption{(a) Fiber B sensor attached to a current collector with five attachment points with two adhesive points, arranged in a straight form. (b) Corresponding measurement results of the thermal changes of the sample from $0$ to $\SI{70}{\degreeCelsius}$ (zero-point calibration at $\SI{20}{\degreeCelsius}$).}
		\label{fig:Anode1}
	\end{figure}  		
	
\begin{figure}[ht!]
		\centering
		\subcaptionbox{\label{fig:AnodeSample2}}{\includegraphics[width=0.9\textwidth]{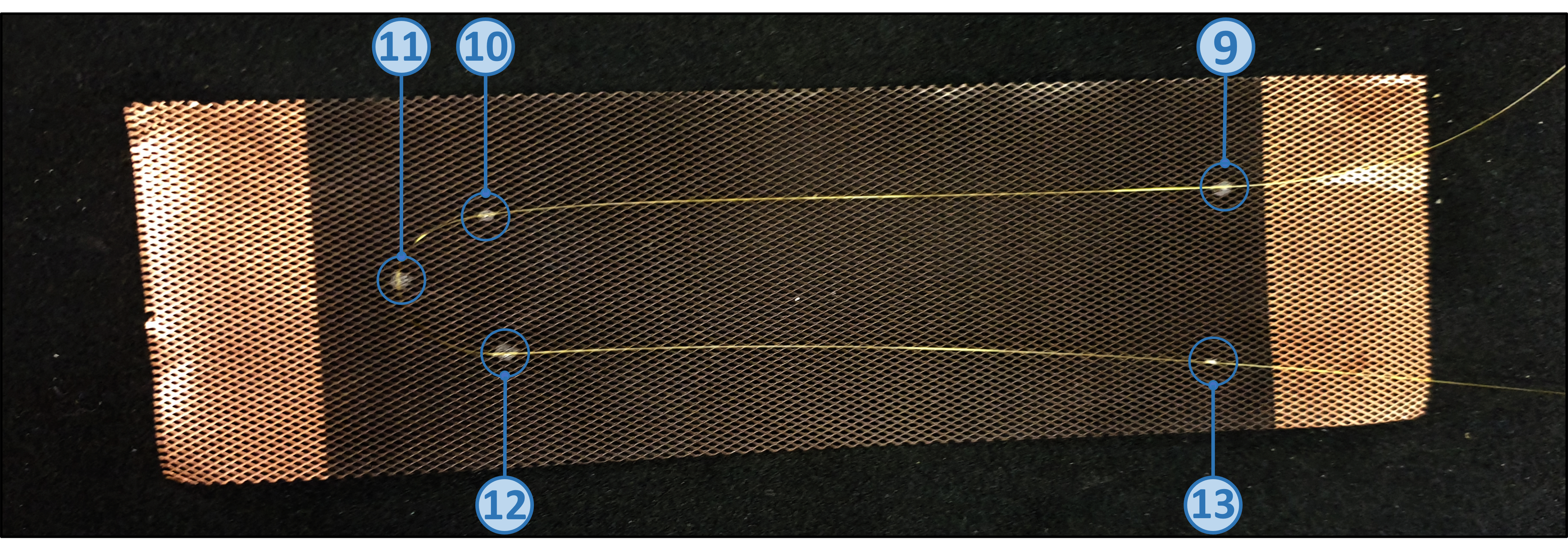}}
		\smallskip
		\subcaptionbox{\label{fig:AnodeSample2_T}}{\includegraphics[width=0.9\textwidth]{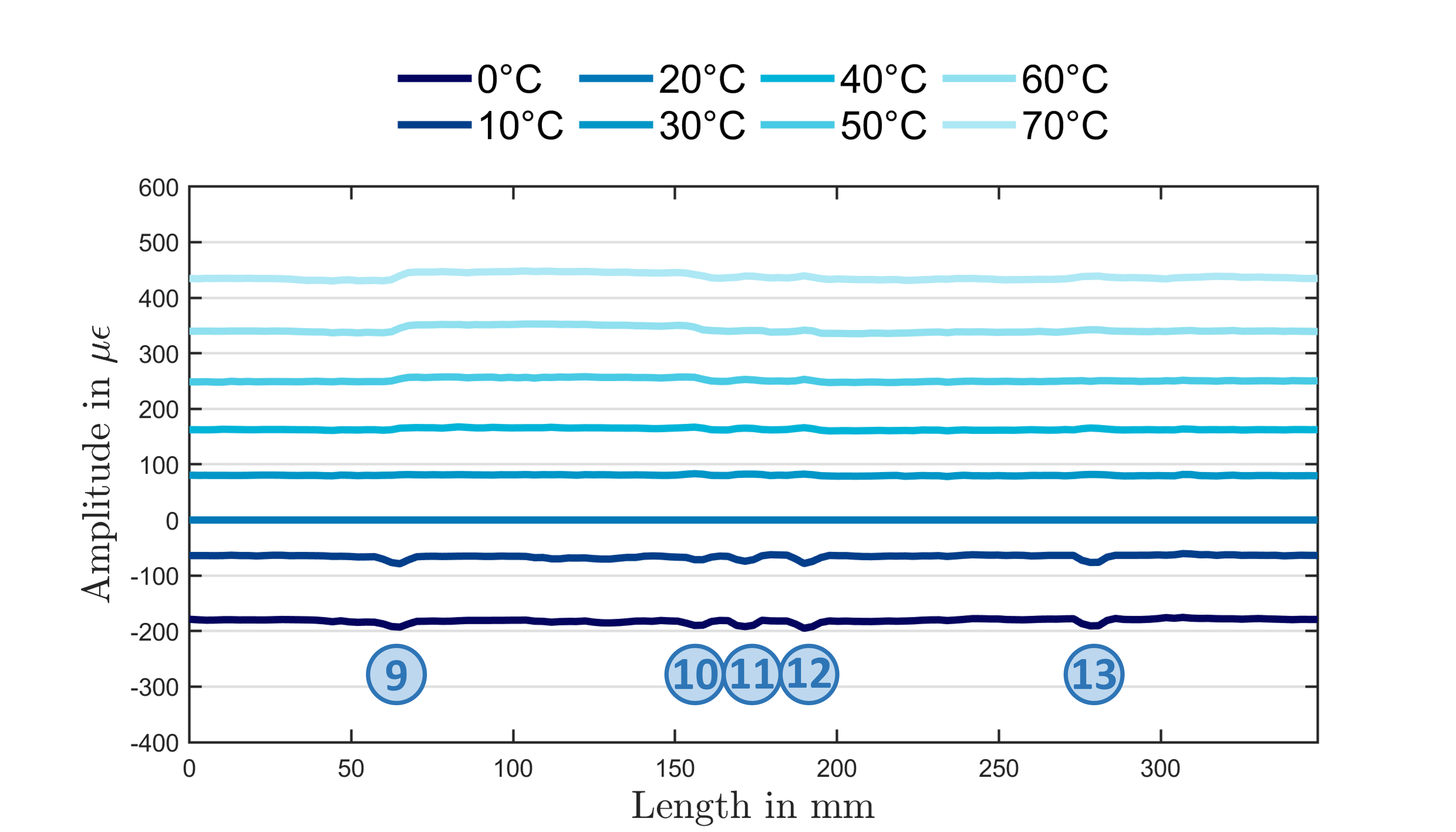}}
		\caption{(a) Fiber B sensor attached to a current collector with five adhesive points, arranged in a U-shape. (b) Corresponding measurement results of the thermal changes of the sample from $0$ to $\SI{70}{\degreeCelsius}$ (zero-point calibration at $\SI{20}{\degreeCelsius}$).}
		\label{fig:Anode2}
	\end{figure}

	Instead of simply inserting the fiber between the outer pouch layer and the cell stack,the fibers can be immobilised on cell components, in order to prevent undesired motion. Introducing adhesive spots to attach the fiber to the electrode might not only impact the battery cell properties, but also the measurement results of the fiber sensor. Here, too, the focus is set on examining how these adhesive points might impact the sensor signal. We prepared various samples of fibers immobilised on expanded copper foil measuring $\SI{160}{\mm}$ x $\SI{60}{\mm}$ (LxW), as depicted in Figure~\ref{fig:AnodeSample1} and Figure~\ref{fig:AnodeSample2}, to analyze the possible effects of the adhesive points on the measurement properties of the fiber.
	
	In the following the temperature-dependent measurement results of the sensors on the samples are shown over the length of the fiber, with taring at $\SI{20}{\degreeCelsius}$. Figure~\ref{fig:Anode1} shows a sample with a straight fiber geometry and two adhesive points ($\circled{7}$ and $\circled{8}$). An u-shaped fiber routing with a radius of $\SI{15}{\mm}$ and five adhesive points ($\circled{9}$, $\circled{10}$, $\circled{11}$, $\circled{12}$ \& $\circled{13}$) is seen in Figure~\ref{fig:Anode2}. 
	
	The temperature measurement results shown in Figure~\ref{fig:AnodeSample1_T}, specifically for the straight geometry and fixation of the fiber at $\circled{7}$ and $\circled{8}$, confirm the assumption that these particular adhesive points slightly influence the measurement outcomes. However, this influence is confined to the bonded areas and does not significantly impact the overall behavior of the fiber. 
	The results exhibit a quasi-constant trend along the fiber length, with deviations arising from the mechanical effects of the adhesive. With increasing temperature, we observe a rise in signal scattering and standard deviation up to  $\SI{10,4}{\micro\epsilon}$ at $\SI{70}{\degreeCelsius}$. This is primarily attributed to negative influences of the climate chamber, as reported earlier. Due to the chamber's heating and the activation of air circulation, the fiber moves repeatedly, distorting the measurement results. However, the standard deviation for the remaining temperature ranges is only $\SI{2,58}{\micro\epsilon}$. Furthermore, the average temperature dependence of $\mi{K}{T} = \SI{8,02}{\frac{\micro\epsilon}{\degreeCelsius}}$ falls within an acceptable range.
		
	The results of the temperature measurement for the u-shaped geometry of the fiber show similar results, see Figure~\ref{fig:AnodeSample2_T}. The relatively small increase in the area of the adhesive spots $\circled{9}$ and $\circled{13}$ in particular visible at $\SI{70}{\degreeCelsius}$ are due to the fact that the bonding spot $\circled{9}$ detached and the fiber had started to move slightly due to the ventilation of the climatic chamber. This scattering of the results is also evident in the standard deviation, which is $\SI{10,9}{\micro\epsilon}$ between $50-\SI{70}{\degreeCelsius}$, i.e. with increased chamber activity, compared to the otherwise average of $\SI{3,46}{\micro\epsilon}$. Although the measurements of the sensors were influenced, temperature measurement results increases nevertheless are proportionally along the fiber. The average temperature dependence results in $\mi{K}{T} = \SI{8,04}{\frac{\micro\epsilon}{\degreeCelsius}}$. Also noticeable here is the absence of any tendency toward measurement inaccuracies in the existing bend between points $\circled{10}$, $\circled{11}$ and $\circled{12}$. Hence, a valid temperature measurement can be performed.

	\section{Conclusion}
	\label{sec:Conclusion}
	
	This work systematically investigated the application of fiber-optic sensors for thermal monitoring in various scenarios, with a primary focus on lithium-ion battery cells. Employing the \acrshort{acr:OFDR} method, the study demonstrated that these sensors can accurately detect and measure spatially resolved temperature changes across a range of conditions, proving their potential for localized thermal anomaly detection. Experiments across multiple setups and fiber types consistently revealed temperature sensitivities within the range $\mi{K}{T} = 8-\SI{11}{\frac{\micro\epsilon}{\degreeCelsius}}$, closely matching values reported in the literature.
	
	\subsection*{Key Findings}
	\begin{enumerate}
		\item \textbf{Temperature Measurement:} 
		Fiber optic sensors showed consistent linear behavior across the tested range ($0$ to $\SI{80}{\degreeCelsius}$) with a resolution of $\SI{2.6}{\mm}$ and a sampling rate of $\SI{1}{\hertz}$. These sensors were effective for real-time, spatially resolved temperature monitoring, crucial for preventing thermal runaways and optimizing battery performance and aging behavior.
		
		\item \textbf{Fiber Integration:} 
		The minimal diameter of the fibers ($155-\SI{250}{\um}$) enabled close placement to or within cells without creating significant stress points, making them highly suitable for integration. Practical tests showed that bending fibers to radii as small as $\SI{15}{\mm}$ did not compromise measurement accuracy or structural integrity. However, tighter bends or deviations from manufacturer specifications could still affect light propagation and need further evaluation.
		
		\item \textbf{Impact of Fixation:} 
		The integration of fibers into pouch cells, which required careful consideration of sealing seams, did not compromise internal sensitivity of the fiber optic sensor. However, adhesive fixation points and sealing seams caused localized mechanical stress, impacting measurement results only in the immediate vicinity of the attachments. Beyond these localized areas, measurements remained reliable and unaffected, emphasizing the potential for embedding fibers into battery systems.		
		
		\item \textbf{Limitations:} 
		Mechanical influences, though avoided in this study, can affect fibers during real-world applications. These influences, including internal battery expansion or external mechanical stress, should be explicitly addressed in future research. The study was limited to thermally induced effects and further investigations are needed to understand and mitigate non-thermal factors affecting sensor performance.
	\end{enumerate}

	Fiber optic sensors exhibit strong potential for improving thermal monitoring of battery systems, offering precise, spatially resolved data critical for safety and performance optimization. Their ability to withstand aggressive chemical environments, small form factor and high sensitivity position them as a versatile tool for battery monitoring, both on the surface and internally.
	
	Future research should explore the long-term reliability of these sensors under mechanical and environmental stresses typical of battery operation. By addressing these challenges, fiber optic sensors could play a transformative role in advancing safer, more efficient and sustainable battery technologies. This work lays the groundwork for integrating these sensors into monitoring devices, paving the way for broader applications in energy storage systems.

	\section*{CRediT authorship contribution statement}
	
	\textbf{Florian Krause}: Conceptualization, Methodology, Investigation, Writing - Original Draft, Writing - Review \& Editing, Visualization. \textbf{Felix Schweizer}: Methodology, Investigation, Visualization. \textbf{Alexandra Burger}: Conceptualization, Methodology, Investigation, Resources, Writing - Review \& Editing, Visualization. \textbf{Franziska Ludewig}: Conceptualization, Writing - Review \& Editing. \textbf{Marcus Knips}: Conceptualization, Writing - Review \& Editing. \textbf{Katharina Quade}: Supervision, Writing - Review \& Editing. \textbf{Andreas Würsig}: Supervision, Funding acquisition. \textbf{Dirk Uwe Sauer}: Supervision, Funding acquisition, Writing - Review \& Editing.
	
	\section*{Acknowledgments}
	
	This work was supported by the Federal Ministry of Education and Research from Germany (BMBF) through the project NUBase under grant numbers 03XP0322C. Furthermore, we express gratitude to Marian Walter for his comments and inspiring scientific discussions.

	\section*{Data availability}
	
	The raw/processed data required to reproduce these findings cannot be shared at this time as the data also forms part of an ongoing study.

	\section*{Disclaimer on the use of AI writing aids}
	
	During the preparation of the manuscript the authors used GPT3.5/OpenAI in order to improve readability and language. After using this tool, the authors reviewed and edited the content as needed and take full responsibility for the content of the publication.

	\bibliographystyle{elsarticle-num}
	\bibliography{Glasfaser_Paper}

\end{document}